\documentclass[aps,
prd,
nofootinbib,
showpacs,
12pt,
superscriptaddress,
preprintnumbers,
floatfix]{revtex4-2}

\usepackage[bookmarks, linktocpage, colorlinks = true, linkcolor = blue, urlcolor  = blue, citecolor = blue, anchorcolor = green, hyperindex = true, hyperfigures]{hyperref}

\usepackage{graphicx} 
\usepackage{dcolumn}
\usepackage{multirow}
\usepackage{amsmath, amssymb}
\usepackage{slashed}
\usepackage[usenames]{color}
\usepackage{float}

\newcommand{\gbar}{\bar{g}}
\newcommand{\Nc}{N_{c}}
\newcommand{\Nf}{N_{f}}
\newcommand{\calD}{\mathcal{D}}
\newcommand{\calM}{\mathcal{M}}
\newcommand{\calP}{\mathcal{P}}

\newcommand{\bkhat}{\hat{\boldsymbol{k}}}
\newcommand{\bl}{\boldsymbol{l}}
\newcommand{\bp}{\boldsymbol{p}}
\newcommand{\bphat}{\hat{\boldsymbol{p}}}
\newcommand{\bq}{\boldsymbol{q}}
\newcommand{\pF}{p_{\mathrm{F}}}
\newcommand{\mD}{m_{\mathrm{D}}}
\newcommand{\mg}{m_{\mathrm{g}}}
\newcommand{\DE}{D^{\mathrm{E}}}
\newcommand{\DM}{D^{\mathrm{M}}}
\newcommand{\OmegaE}{\Omega}
\newcommand{\Vsing}{V^{\mathrm{(s)}}}
\newcommand{\Vtrip}{V^{\mathrm{(t)}}}

\begin{document}

\preprint{N3AS-25-016, RIKEN-iTHEMS-Report-25}
\title{Renormalization group analysis of color superconductivity revisited}

\author{Yuki~Fujimoto}
\affiliation{Department of Physics, Niigata University, Ikarashi, Niigata 950-2181, Japan}
\affiliation{Department of Physics, University of California, Berkeley, CA 94720, USA}
\affiliation{RIKEN Center for Interdisciplinary Theoretical and Mathematical Sciences (iTHEMS), RIKEN, Wako 351-0198, Japan}

\date{\today}

\begin{abstract}
    Color superconductivity in cold, dense quark matter is a key feature of the QCD phase diagram, whose present theoretical understanding relies predominantly on weak-coupling calculations.
    In this work, we revisit the evaluation of the color-superconducting gap using a renormalization group (RG) framework formulated in effective theory near the Fermi surface.
    By incorporating quark self-energy corrections into the RG equation, we reproduce the known weak-coupling results from the gap equation at the same perturbative order at $O(g^0)$.
    Within the RG approach, the angular momentum structure of the pairing channel becomes more transparent, allowing us to examine the size of the gap for various pairing patterns.
    We also compare our results with recent lattice QCD calculations at finite isospin density.
    Finally, we argue that the RG method potentially offers a simpler and more systematic route to higher-order computations of the gap, which are of order $O(g)$ and thus quantitatively important.
\end{abstract}

\maketitle

\section{Introduction}

Cold quark matter at finite baryon density is expected to be a color superconductor, induced by a diquark condensate (see Ref.~\cite{Alford:2007xm} and reference therein).
Quark Cooper pairing is inevitable because there is always an attractive channel in quark-quark interaction;
whenever there is an attraction between particles on the Fermi surface, it leads to BCS instability and formation of Cooper pairs.
Onset of BCS instability can be seen transparently by using the renormalization group (RG) approach to the Fermi surface, which was pioneered in early 1990s~\cite{BenfattoGallavotti1990,Benfatto:1990zz, Benfatto:1996ng, Feldman1990PerturbationTF, Feldman1991TheFO, Feldman1992AnIV, Feldman:1993ck, SHANKAR1991530, Shankar:1993pf, Polchinski:1992ed}.
The same RG technology can be applied to relativistic quarks in cold and dense matter to study color superconductivity~\cite{Evans:1998ek,Evans:1998nf,Schafer:1998na,Son:1998uk,Hsu:1999mp}.
In QCD at finite baryon density, there is a complication related to a statically unscreened chromomagnetic gluon, as first pointed out in Ref.~\cite{Son:1998uk};
this long-range magnetic interaction gives rise to the amplification in the pairing gap compared to the short-range interaction case.

The current understanding of color superconductivity is grounded on the analysis at large quark chemical potential $\mu$, where QCD becomes weakly coupled so that one can calculate the color-superconducting gap $\Delta$ rigorously in perturbation theory.
The pairing gap can either be obtained by solving the gap equation derived in perturbation theory~\cite{Son:1998uk, Pisarski:1998nh, Pisarski:1999av,Pisarski:1999bf, Pisarski:1999tv, Schafer:1999jg, Hong:1998tn, Hong:1999ru,Hong:1999fh, Wang:2001aq} or by reading out the pairing singularity in the fully renormalized two-particle vertex function~\cite{Brown:1999aq, Brown:1999yd, Brown:2000eh}.
Currently, perturbative series of the logarithm of the pairing gap $\ln (\Delta / \mu)$ has been evaluated up to the next-to-leading order at $O(g^0)$, where $g$ is the QCD coupling constant.
The RG approach allows us to find the lowest-order term in the perturbative series of $\ln (\Delta / \mu)$;
historically, at about the same time as the gap equation approach~\cite{Pisarski:1998nh}, it was discovered using the RG approach that the lowest-order term scales as $1/g$, not $1/g^2$~\cite{Son:1998uk} (see also Ref.~\cite{Barrois:1979pv} for the earlier perception on this scaling).

Recently, there is a revived interest in color superconductivity, or quark Cooper pairing more in general.
This is especially in connection with the recent progress in astrophysics and first-principles calculation of QCD, which are specifically lattice QCD and perturbative QCD in this context.
On the astrophysical application side, in the recent work~\cite{Kurkela:2024xfh}, authors constrained the magnitude of the color-superconducting gap from observational data of neutron stars relying on the rigorous interpolation method between the two ab initio calculations at low and high densities~\cite{Komoltsev:2021jzg}.

On the first-principles calculation side, there were remarkable advances in the lattice-QCD calculation of thermodynamic quantities at finite \emph{isospin} chemical potential $\mu_I$~\cite{Abbott:2023coj,Abbott:2024vhj}.
QCD at finite $\mu_I$ is free from the sign problem, and thus amenable to Monte Carlo calculation on the lattice.
In the latest lattice calculation in the continuum limit~\cite{Abbott:2024vhj}, the equation of state was obtained up to large chemical potential, $\mu_I \simeq 3 \,\text{GeV}$, at which QCD becomes weakly coupled, so direct comparison with the weak-coupling calculation is possible~\cite{Fujimoto:2023mvc,Fujimoto:2024pcd,Fukushima:2024gmp}.
In the weak-coupling result, the effect of the pairing gap is substantial because QCD at large $\mu_I$ exhibits quark superfluidity, and its gap is significantly larger compared to that of color superconductivity~\cite{Son:2000xc}.
We find qualitative agreement between the lattice-QCD and weak-coupling calculations, but quantitatively, the pairing gap contribution in the weak-coupling calculation seems to be overestimated compared to the latest lattice calculation~\cite{Fujimoto:2024pcd,Fukushima:2024gmp}.

In this work, we revisit the RG analysis of color superconductivity at weak coupling.
We demonstrate quantitatively that the RG analysis can also lead to the weak-coupling expression of the gap up to the currently known order $O(g^0)$.
In particular, we include the self-energy correction into the RG equation, which can be incorporated by using the resummed quark propagator in hard dense loop (HDL) effective theory.
We also systematically re-derive the constant term in the beta function arising from the long-range interaction, which was first pointed out by Son~\cite{Son:1998uk}.

Although the current work does not push the limit of the existing calculation, there are several advantages of the RG approach over the other approaches to color superconductivity at weak coupling.
The RG equation decouples in each spin and orbital angular momentum channel, so the angular momentum state in which the pairing occurs becomes clear.
This is related to the accompanying paper~\cite{Fujimoto:2025liq} in which we classify the possible spin and orbital angular momentum channels using the term symbols as in the nonrelativistic case.
This is possible because of the loss of Lorentz invariance at finite density.
Another advantage is that the RG approach provides good insight into  higher-order calculation in perturbation theory.
All the perturbative corrections in the effective theory near the Fermi surface, for which we derive the RG equation, can be classified by whether they are relevant, marginal, or irrelevant, on top of the ordinary expansion in terms of the coupling constant.
The combination of these two hierarchies simplifies the organization of the higher-order corrections.

The paper is organized as follows.
In Sec.~\ref{sec:Fermi}, we review the RG approach near the Fermi surface.
The readers who are already familiar with this can skip to the next section.
In Sec.~\ref{sec:RGeq}, we derive the RG equation: First, in Sec.~\ref{sec:oneloop}, we calculate the one-loop beta function, which was already obtained in Refs.~\cite{Evans:1998ek,Evans:1998nf,Schafer:1998na}.
Then, in Sec.~\ref{sec:self}, we derive the self-energy correction in the RG equation, which is new in this work.
In Sec.~\ref{sec:tree}, we re-derive Son's beta function in Ref.~\cite{Son:1998uk} systematically from a renormalization of the tree-level amplitude.
Finally, we collect the all the elements in the RG equation in Sec.~\ref{sec:fullRG}.
In Sec.~\ref{sec:solution}, we solve the RG equation and obtain the pairing gap from the singularity in the solution.
In Sec.~\ref{sec:gapeq}, we compare our solution with the solution of the gap equation in the preceding works.
In Sec.~\ref{sec:isospin}, we discuss the superfluid gap in QCD at finite isospin density and contrast our weak-coupling calculation with the lattice-QCD data, which is the only cross check we can perform at the moment.
In Sec.~\ref{sec:discussions}, we discuss several effects related to the gap.
Finally, we conclude the paper in Sec.~\ref{sec:summary}.

\section{Renormalization group approach near the Fermi surface}
\label{sec:Fermi}

To make the paper self-contained and to introduce the formulas necessary later in the text, we review the renormalization group (RG) approach to Landau Fermi liquid~\cite{Shankar:1993pf,Polchinski:1992ed}.
The readers already familiar with the context can skip to the next section.

Let us analyze interacting massless quark excitations near the Fermi surface using RG.
We consider the effective theory with the kinetic term of the action as
\begin{align}
    S_0 = \int_{|\epsilon_p| < \Lambda}\frac{d^4 p}{(2\pi)^4} \bar{\psi}(p) (p_0 - \epsilon_p) \psi(p)\,,
\end{align}
where $\epsilon_p = |\bp| - \mu$ is the energy relative to the Fermi surface and $p_\mu = (p_0, \bp)$ is the Minkowskian four-momentum.
Note that the chemical potential $\mu$ and the Fermi momentum $\pF$ matches for the massless fermion, i.e., $\mu = \pF$.
We limit the effective theory to contain only the quarks located in a thin shell surrounding the Fermi surface with a cutoff $\Lambda \ll \mu$, i.e., $|p_0| < \Lambda$ and $|\epsilon_p| < \Lambda$.
As the cutoff scales to zero, the momentum scales toward the Fermi surface.

Let us apply the RG transformation to the quartic coupling function $G$ in the interaction term of the following form:
\begin{align}
    S_{\mathrm{I}}
    &= \frac{1}{4} \left(\prod_{i=1}^4 \int_{|\epsilon_p| < \Lambda} \frac{d^4 p_i}{(2\pi)^4} \right) \mathcal{L}_{\mathrm{I}}\notag \\
    &\qquad \times (2\pi)^4 \delta^{(4)}(p_1 + p_2 - p_3 - p_4) \,,
    \label{eq:SI}\\
    \mathcal{L}_{\mathrm{I}} &= G_{\mu\nu}(p_1, p_2, p_3, p_4) \bar{\psi}(p_3) \Gamma^\mu \psi(p_1) \bar{\psi}(p_4) \Gamma^\nu \psi(p_2) \,,
\end{align}
where $\Gamma^\mu$ refers to an arbitrary combination of the Dirac matrices.
The partition function is
\begin{align}
    \mathcal{Z} = \mathcal{N}\int_{l < \Lambda} \!\!\!\!\!\! \calD \bar{\psi}(l)  \calD \psi(l)\, e^{i S_0[\psi] + i S_{\mathrm{I}}[\psi]}\,,
    \label{eq:Z1}
\end{align}
where $\mathcal{N}$ is the normalization constant and the four-momentum $l$, whose spatial component is measured relative to the Fermi momentum $\mu$, is defined as
\begin{align}
    l_\mu = (\omega, \bl) \equiv (p_0, \epsilon_p \bphat)\,,
\end{align}
and $l < \Lambda$ in the integration means that the integral is taken over the momentum $|\omega| < \Lambda$ and $|\bl|< \Lambda$.
The Lorentz invariance is lost anyway because of the medium $\mu \neq 0$.

In Eq.~\eqref{eq:SI}, the argument of the delta function in the spatial direction is
\begin{align}
    \bp_1 + \bp_2 - \bp_3 - \bp_4
    = \mu (\bphat_1 + \bphat_2 - \bphat_3 - \bphat_4)
    + \bl_1 + \bl_2 - \bl_3 - \bl_4\,.
\end{align}
We limit ourselves to the kinematics relevant for BCS instability
\begin{align}
    \bphat_1 = - \bphat_2\,,\qquad
    \bphat_3 = - \bphat_4\,,
    \label{eq:kinematics}
\end{align}
so that $\bphat_1 + \bphat_2 - \bphat_3 - \bphat_4 = 0$.
This allows us to rewrite Eq.~\eqref{eq:SI} by $l_i$ as
\begin{align}
    S_{\mathrm{I}}
    &= \frac{1}{4} \left(\prod_{i=1}^4 \int_{l < \Lambda} \frac{d^4 l_i}{(2\pi)^4} \right) \mathcal{L}_{\mathrm{I}}\notag \\
    &\qquad \times (2\pi)^4 \delta^{(4)}(l_1 + l_2 - l_3 - l_4) \,,
    \label{eq:SIl}\\
    \mathcal{L}_{\mathrm{I}} &= G_{\mu\nu}(l_1, l_2, l_3, l_4) \bar{\psi}(l_3) \Gamma^\mu \psi(l_1) \bar{\psi}(l_4) \Gamma^\nu \psi(l_2) \,,
    \label{eq:LIl}
\end{align}
As we will see below, outside this kinematic range~\eqref{eq:kinematics}, this interaction term in the effective theory becomes irrelevant under the RG transformation.

\begin{figure}
    \centering
        \includegraphics[width=0.35\columnwidth]{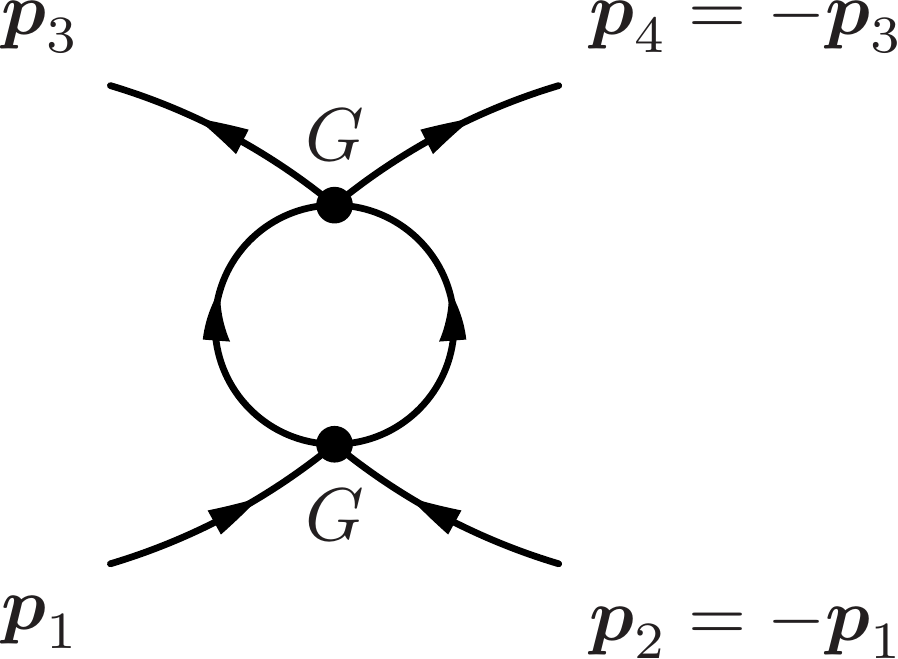}\\
    \caption{The skeleton one-loop diagram that renormalizes the quartic coupling and contributes to the effective theory as marginally relevant.}
    \label{fig:BCS}
\end{figure}
Let us apply the RG transformation to the action, and derive the RG equation for the coupling function.
We split the field variable into the slow mode $\psi_<$ and the fast mode $\psi_>$
\begin{align}
    \psi(l) &= \psi_<(l) \theta(e^{-t} \Lambda - |\bl|) \theta(e^{-t} \Lambda - |\omega|) \notag\\
    &\quad+ \psi_>(l) \theta(|\bl| - e^{-t} \Lambda) \theta(|\omega| - e^{-t} \Lambda)\,,
\end{align}
where $t>0$, and integrate out the fast mode
\begin{align}
    \mathcal{Z}
    = \mathcal{N}' \int_{l < e^{-t} \Lambda} \!\!\!\!\!\!\!\!\!\!\!\!\! \calD \bar{\psi}_<(l)  \calD{\psi_<(l)}\, e^{i S_0[\psi_<] + i S_{\mathrm{I}}^{<}[\psi_<]}\,,
    \label{eq:Z2}
\end{align}
where $\mathcal{N}'$ is the normalization constant after the mode elimination.
The effective action after integrating out the fast mode is
\begin{align}
    e^{i S_{\mathrm{I}}^<[\psi_<]} &= \langle e^{i S_{\mathrm{I}}[\psi_<, \psi_>]}\rangle_{>0}\,,\notag \\
    &= e^{\left[ i\langle S_{\mathrm{I}}\rangle_{>0} - \frac12 \left(\langle S_{\mathrm{I}}^2\rangle_{>0} - \langle S_{\mathrm{I}}\rangle_{>0}^2 \right) + \cdots\right]}\,,
\end{align}
where $\langle \cdot \rangle_{0>}$ refers to the averages with respect to the fast modes with action $S_0$
\begin{align}
    \langle \mathcal{O}[\psi_<, \psi_>] \rangle_{>0} \equiv \frac{\displaystyle \int_{e^{-t} \Lambda \le l < \Lambda} \!\!\!\!\!\!\!\!\!\!\!\!\!\!\!\!\!\! \calD \bar{\psi}_>(l)  \calD{\psi_>(l)}\, e^{i S_0[\psi_>]} \mathcal{O}[\psi_<, \psi_>]}
    {\displaystyle \int_{e^{-t} \Lambda \le l < \Lambda} \!\!\!\!\!\!\!\!\!\!\!\!\!\!\!\!\!\! \calD \bar{\psi}_>(l)  \calD{\psi_>(l)}\, e^{i S_0[\psi_>]}}\,.
\end{align}
The renormalized effective can be decomposed into the tree-level piece $S_{\mathrm{I,tree}}^<$ and the one-loop corrections $\delta S_{\mathrm{I}}^<$ arising from the cumulant expansion above:
\begin{align}
    S_{\mathrm{I}}^<[\psi_<] &= S_{\mathrm{I,tree}}^<[\psi_<] + \delta S_{\mathrm{I}}^<[\psi_<] + \cdots \,,\\
    S_{\mathrm{I,tree}}^<[\psi_<] &= \langle S_{\mathrm{I}}\rangle_{>0} \,,\label{eq:SIlesstree}\\
    \delta S_{\mathrm{I}}^<[\psi_<] &= \frac{i}2 \left(\langle S_{\mathrm{I}}^2\rangle_{>0} - \langle S_{\mathrm{I}}\rangle_{>0}^2 \right)\,.
\end{align}
The Feynman diagram for the one-loop correction is shown in Fig.~\ref{fig:BCS}.
This is the only marginally relevant contribution, and all the other one-loop and higher-order diagrams turn out to be irrelevant~\cite{Shankar:1993pf}.

We have integrated out the fast mode, i.e., reduced the cutoff as $\Lambda \to e^{-t} \Lambda$.
Now, to compare this equation with the original expression of the partition function~\eqref{eq:Z1}, we rescale the momentum as
\begin{align}
\begin{split}
    l_\mu' &= e^t l_\mu\,,\\
    \psi'(l') &= e^{-3t/2} \psi_<(l' e^{-t} ) = e^{-3t/2} \psi_<(l)\,.
\end{split}
    \label{eq:rescaled}
\end{align}
Rescaled momenta now go all the way to $\Lambda$ again, so the direct comparison with Eq.~\eqref{eq:Z1} is now possible.
The $t$-scaling of the fermion field $e^{3t/2}$ is chosen so that the kinetic term $S_0$ has the same coefficient as $S_0$ before the RG transformation.
In terms the rescaled variables, the partition function is
\begin{align}
    \mathcal{Z}
    = \mathcal{N}' \int_{l' < \Lambda} \!\!\!\!\!\! \calD \bar{\psi}'(l')  \calD \psi'(l')\, e^{i S_0[\psi'] + i S_{\mathrm{I}}'[\psi']}\,.
    \label{eq:Zprime}
\end{align}
This procedure completes the RG transformation.

Now, let us focus on the RG transformation of the tree-level amplitude.
As defined in Eq.~\eqref{eq:SIlesstree}, $S_{\mathrm{I,tree}}^<$ reads
\begin{align}
    S_{\mathrm{I,tree}}^<[\psi_<]
    &= \frac{1}{4} \left(\prod_{i=1}^4 \int_{l < e^{-t} \Lambda} \!\! \frac{d^4 l_i}{(2\pi)^4} \right) \mathcal{L}_{\mathrm{I,tree}}^< \notag\\
    &\qquad \times(2\pi)^4 \delta^{(4)}(l_1 + l_2 - l_3 - l_4) \,,\\
    \mathcal{L}_{\mathrm{I,tree}}^< &= G_{\mu\nu}(l_1, l_2, l_3, l_4)
    \bar{\psi}_<(l_3) \Gamma^\mu \psi_<(l_1) \bar{\psi}_<(l_4) \Gamma^\nu \psi_<(l_2) \,.
\end{align}
One can rewrite this in terms of the rescaled variables~\eqref{eq:rescaled}
\begin{align}
    S_{\mathrm{I,tree}}'[\psi']
    &= e^{-2t} \frac{1}{4} \left(\prod_{i=1}^4 \int_{l' < \Lambda} \!\! \frac{d^4 l_i'}{(2\pi)^4} \right) \mathcal{L}_{\mathrm{I,tree}}' \notag\\
    &\qquad \times (2\pi)^4 \delta^{(4)}\left(e^{-t} l_1' + e^{-t} l_2' - e^{-t} l_3' - e^{-t} l_4'\right) \,,
    \label{eq:SItree}\\
    \mathcal{L}_{\mathrm{I,tree}}' &= G_{\mu\nu}(e^{-t} l_1', e^{-t} l_2', e^{-t} l_3', e^{-t} l_4')
    \bar{\psi}'(l_3') \Gamma^\mu \psi'(l_1') \bar{\psi}'(l_4') \Gamma^\nu \psi'(l_2') \,,
    \label{eq:LItree}
\end{align}
Due to the factor $e^{-2t}$ in front, this term is irrelevant unless the  delta function scales as
\begin{align}
    \delta^{(4)}\left(e^{-t} l_1' + e^{-t} l_2' - e^{-t} l_3' - e^{-t} l_4'\right)
    = e^{2t}\delta^{(4)}(l_1' + l_2' - l_3' - l_4')
\end{align} 
This is only possible when the directions of the momenta fulfill the condition Eq.~\eqref{eq:kinematics}.
Therefore, the quartic coupling remains marginal under the RG transformation only when $\bphat_1 + \bphat_2 - \bphat_3 - \bphat_4 = 0$, and it is irrelevant otherwise.

Likewise, one can schematically write the one-loop correction piece $\delta S_{\mathrm{I}}'$ as
\begin{align}
    \delta S_{\mathrm{I}}'[\psi']
    &= \frac{1}{4} \left(\prod_{i=1}^4 \int_{l' < \Lambda} \!\! \frac{d^4 l_i'}{(2\pi)^4} \right) \delta \mathcal{L}_{\mathrm{I}}' \notag\\
    &\qquad \times (2\pi)^4 \delta^{(4)}\left(l_1' + l_2' - l_3' - l_4'\right) \,,
    \label{eq:SIloop}\\
    \delta \mathcal{L}_{\mathrm{I}}' &= \delta G_{\mu\nu} \, \bar{\psi}'(l_3') \Gamma^\mu \psi'(l_1') \bar{\psi}'(l_4') \Gamma^\nu \psi'(l_2') \,,
    \label{eq:LIloop}
\end{align}
The detailed computation will be shown in the next section.

By comparing Eqs.~(\ref{eq:Z1}, \ref{eq:Zprime}) and using Eqs.~(\ref{eq:SIl}, \ref{eq:LIl}, \ref{eq:SItree}, \ref{eq:LItree}, \ref{eq:SIloop}, \ref{eq:LIloop}), one finds the RG equation for the coupling function
\begin{align}
    G(l_1, l_2, l_3, l_4) = G(e^{-t} l_1, e^{-t} l_2, e^{-t} l_3, e^{-t} l_4) + \delta G\,,
\end{align}
where we suppressed the Lorentz indices $\mu\nu$ in $G_{\mu\nu}$.
In terms of the differential equation
\begin{align}
    \frac{dG}{dt} = \beta_{\mathrm{tree}} + \delta \beta\,,
\end{align}
where $\beta_{\mathrm{tree}} = dG(e^{-t} l_1, e^{-t} l_2, e^{-t} l_3, e^{-t} l_4)/dt$ is from the renormalization of the tree-level amplitude, and $\delta \beta$ arises from the one-loop correction to the coupling function $\delta G$.
The tree-level beta function $\beta_{\mathrm{tree}}$ vanishes for a coupling function that is insensitive to the soft mode near the Fermi surface, e.g., a coupling function of a constant value.
We will see below that this is not the case for the long-range interaction.

\section{Renormalization group equation in dense QCD}
\label{sec:RGeq}

Now, we consider the renormalization of the coupling function in the four-fermion interaction in an effective theory near the Fermi surface of the form\footnote{Note that the sign of $G^i$ is opposite to the definition in preceding papers~\cite{Evans:1998ek, Evans:1998nf,Schafer:1998na,Son:1998uk}.
The overall sign is chosen so that $G^0$ and $G^i$ are equal to the electric and gluon propagators, respectively, when considering the matching condition with the perturbation theory in Eq.~\eqref{eq:matching}.}
\begin{align}
    \mathcal{L}_{\mathrm{I}} = -G^0 (\bar{\psi}\gamma^0 \psi)^2 + G^i (\bar{\psi}\gamma^i \psi)^2\,.
\end{align}
We only consider this interaction term near the Fermi surface, i.e., $|\bp_i| \simeq \mu$ and $\big||\bp_i| - \mu\big| \lesssim \Lambda$, and under the kinematic constraint \eqref{eq:kinematics}, in which the incoming momenta $\bp_1$ and $\bp_2$ and the outgoing momenta $\bp_3$ and $\bp_4$ fulfill the relations $\bp_1 = - \bp_2$ and $\bp_3 = -\bp_4$, respectively.
So, Cooper pairs have zero net momentum;
in this paper, we limit ourselves to considering the non-crystalline color superconductivity.
Without loss of generality, the scattering angle can be defined as $\bphat_1 \cdot \bphat_3 = \cos\theta$.
The tree-level amplitude calculated from this interaction term is
\begin{align}
    i\calM_{\mathrm{tree}} &= iG_{\mu\nu;\lambda_1 \lambda_2} (\bphat_1, \bphat_3)
    \bar{u}_{\lambda_3}(p_3) \gamma^\mu u_{\lambda_1}(p_1) \bar{u}_{\lambda_4}(p_4) \gamma^\nu u_{\lambda_2}(p_2) \,, \label{eq:tree}\\
    G_{\mu\nu;\lambda_1 \lambda_2} &= -G^0_{\lambda_1 \lambda_2} \delta_{\mu 0} \delta_{\nu 0} + G^i_{\lambda_1 \lambda_2} \delta_{\mu i} \delta_{\nu j} \delta_{ij}\,,
    \label{eq:Gmunu}
\end{align}
where $\lambda_i$ is the helicity of the $i$-th particle.
The helicity can take values $\lambda=\pm \frac12$, and we abbreviate them as $\lambda=\pm$.
The helicity conservation requires $\lambda_1 = \lambda_3$ and $\lambda_2 = \lambda_4$, so the helicity structure of this amplitude is completely specified only by $\lambda_1$ and $\lambda_2$.

\begin{figure}
    \centering
    \includegraphics[width=0.75\columnwidth]{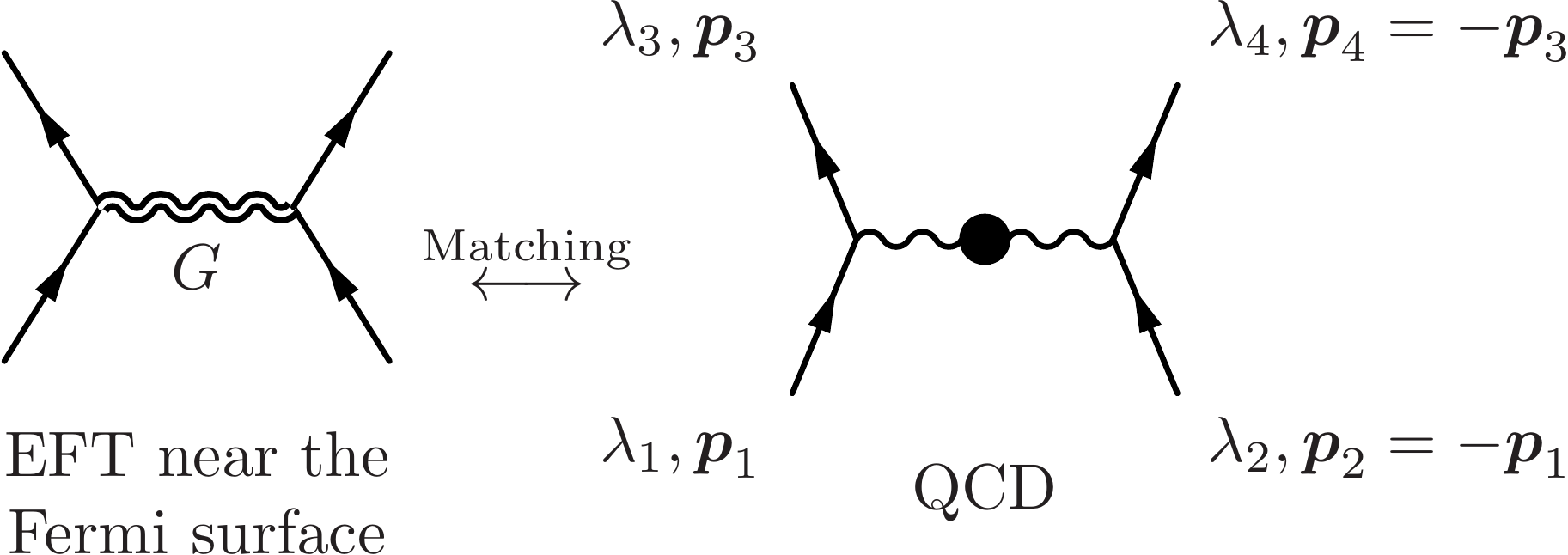}
    \caption{Matching condition of the interaction term $G$ in the effective theory near the Fermi surface to the tree-level amplitude in the HDL effective theory of QCD (right).
    We keep track of the Dirac structure by denoting the interaction $G$ in the effective theory near the Fermi surface by the double-wiggly line, instead of contracting it to a single point.
    The heavy dot in QCD sector indicates the HDL-resummed propagator.}
    \label{fig:matching}
\end{figure}
Matching condition, as shown in Fig.~\ref{fig:matching}, with the tree-level amplitude in perturbation theory, which is the HDL effective theory of QCD in specific, yields
\begin{align}
    G^0 = c_R g^2 \DE\,,\qquad
    G^i = c_R g^2 \DM\,,
    \label{eq:matching}
\end{align}
where $\DE$ and $\DM$ are the HDL-resummed chromoelectric and magnetic propagators
\begin{align}
    D^{\mathrm{E/M}} = - \frac{1}{q^2 - \Pi^{\mathrm{E/M}}}\,,
    \label{eq:gluonprop}
\end{align}
where $q = p_1 - p_3$ and $|\bq| = \sqrt{2\mu^2(1-\cos\theta)}$.
The color factor $c_R$ is
\begin{align}
    c_R = \begin{cases}
        -\frac23 & (\text{color }\boldsymbol{\bar{3}}\text{ channel})\\
        \frac13 & (\text{color }\boldsymbol{6}\text{ channel})\,.
    \end{cases}
    \label{eq:cR}
\end{align}
The HDL resummation is necessary as BCS instability is dominated by the soft gluon interaction in the momentum range $|\bq| \ll \mu$.
In the limit $q_0 \ll |\bq| \to 0$ and to leading order in perturbation theory, the electric (longitudinal) / magnetic (transverse) part of the gluon self-energy $\Pi^{\mathrm{E/M}}$ can be expanded as~\cite{Bellac:2011kqa,Laine:2016hma}
\begin{align}
    \Pi^{\mathrm{E}} &\simeq \mD^2\,,\label{eq:PiE}\\
    \Pi^{\mathrm{M}} &\simeq -i \frac{\pi}4 \mD^2\frac{q_0}{|\bq|}\,,
    \label{eq:PiM}
\end{align}
where $\mD^2 = N_f g^2 \mu^2 / (2\pi^2)$ is Debye screening mass.
The magnetic interaction remains unscreened statically, but there is a dynamical screening with a frequency-dependent cutoff $q_c = (\pi \mD^2 q_0 /4)^{1/3}$.

Before proceeding to the derivation of beta functions, let us comment on the gauge invariance of the RG equation and the pairing gap.
The gauge-dependent term in the propagator is proportional to $q^\mu q^\nu$.
In the present case, because all the quarks are on shell in Eq.~\eqref{eq:tree}, the spinors should respect the current conservation: $q_\mu J^\mu =0$, where the quark current is $J^\mu = \bar{u}_{\lambda_3}(p_3) \gamma^\mu u_{\lambda_1}(p_1)$.
Therefore, from the gauge invariance of the matching condition, the resulting gap in the effective theory also turns out to be gauge invariant.
Related to this, precisely speaking, the projection operator for the spatial part must be
\begin{align}
    P_{\mu\nu}^{\mathrm{T}} = \delta_{\mu i} \delta_{\nu j} (\delta_{ij} - \hat{q}_i \hat{q}_j)\,,
\end{align}
instead of $\delta_{\mu i} \delta_{\nu j} \delta_{ij}$, which was used in Eq.~\eqref{eq:Gmunu}.
Nevertheless, Eq.~\eqref{eq:Gmunu} is correct because the spinors satisfy the relation $\hat{q}_i \bar{u}_{\lambda_3}(p_3) \gamma^i u_{\lambda_1}(p_1) = 0$ in this kinematics.

\subsection{One-loop beta function and the helicity amplitude}
\label{sec:oneloop}

Applying the RG transformation, the one-loop correction to this amplitude $\delta \calM$ is
\begin{align}
    \label{eq:deltaM}
    i \delta\calM_{\lambda_1 \lambda_2} &= i \delta G_{\mu\nu;\lambda_1 \lambda_2}(\bphat_1, \bphat_3)
    \bar{u}_{\lambda_3}(p_3) \gamma^\mu u_{\lambda_1}(p_1) \bar{u}_{\lambda_4}(p_4) \gamma^\nu u_{\lambda_2}(p_2) \,,\\
    \delta G_{\mu\nu;\lambda_1 \lambda_2} &= -\delta G^0_{\lambda_1 \lambda_2} \delta_{\mu 0} \delta_{\nu 0} + \delta G^i_{\lambda_1 \lambda_2} \delta_{\mu i} \delta_{\nu j} \delta_{ij}\,.
\end{align}
\begin{figure}
    \centering
    \includegraphics[width=0.45\columnwidth]{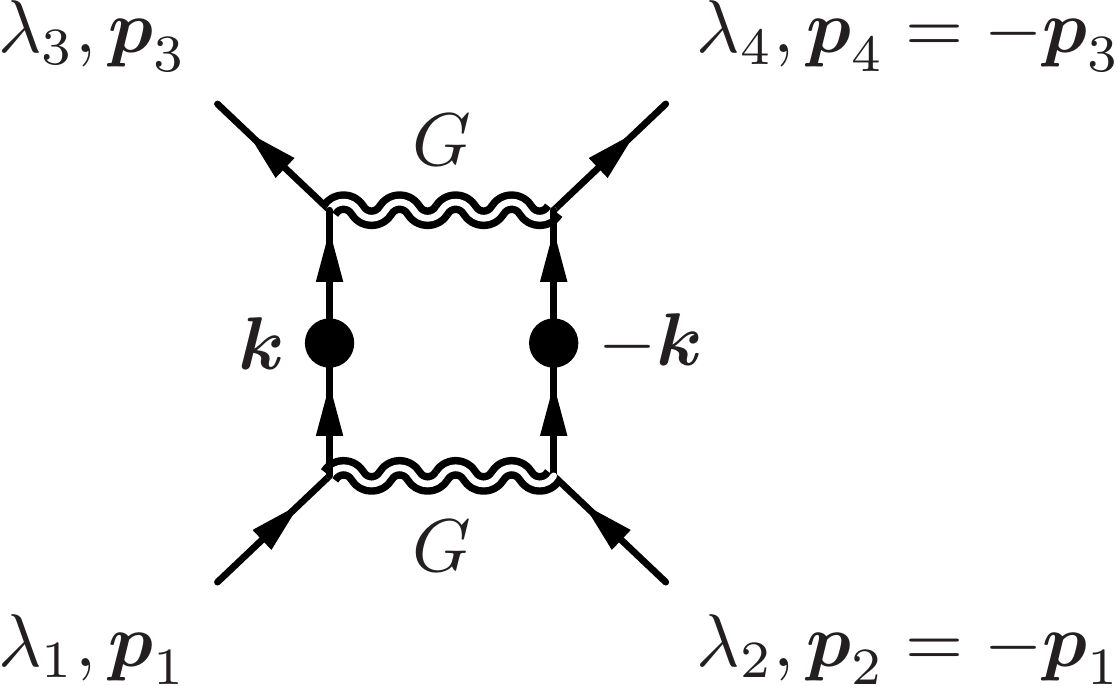}
    \caption{One-loop correction to the beta function.
    The heavy dots indicate the HDL-resummed propagators.}
    \label{fig:ladder}
\end{figure}
The one-loop correction arises from the ladder diagram shown in Fig.~\ref{fig:ladder}.
It reads
\begin{align}
    i \delta \calM_{\lambda_1 \lambda_2}
    &= \int_{d\Lambda} \frac{d^4 k}{(2\pi)^4} \bigg\{
    \left[i G_{\mu\nu;\lambda_1 \lambda_2}(\bphat_1, \bkhat) \right] \left[i G_{\rho\sigma;\lambda_1 \lambda_2}(\bkhat, \bphat_3) \right] \notag\\
    & \qquad \qquad \qquad \times
    \bar{u}_{\lambda_3}(p_3) \gamma^\rho \left(\frac{i\gamma^0 \calP^+_{k}}{k_0 - \epsilon_k}\right) \gamma^\mu u_{\lambda_1}(p_1) \notag \\
    & \qquad \qquad \qquad \times \bar{u}_{\lambda_4}(p_4) \gamma^{\sigma}\left(\frac{i\gamma^0 \calP^+_{-k}}{-k_0 - \epsilon_k}\right) \gamma^\nu u_{\lambda_2}(p_2) \bigg\}\,,
\end{align}
where $d\Lambda$ is the shorthand notation for $e^{-t} \Lambda< |\epsilon_k| < \Lambda$, and 
$\calP_{k}^{\pm} \equiv (1 \pm \gamma_0 \gamma^i \hat{k}_i) / 2$ is the energy projector.
Here, we do not use the resummed propagator for quarks, and will include them in the next subsection.
After straightforward algebra, one can decompose as
\begin{align}
    \label{eq:deltaMIT}
    \delta \calM_{\lambda_1 \lambda_2} = \mathcal{I} \mathcal{T}_{\lambda_1 \lambda_2}\,,
\end{align}
where $\mathcal{I}$ and $\mathcal{T}_{\lambda_1 \lambda_2}$ are defined as
\begin{align}
    \label{eq:calI}
    \mathcal{I} &\equiv -\frac14 \int_{-\infty}^{\infty} \frac{dk_0}{2\pi} \int_{d\Lambda} \frac{k^2 \, d k}{2\pi^2} \,
    \frac{1}{k_0^2 - \epsilon_k^2}\,, \\
    \mathcal{T}_{\lambda_1 \lambda_2}
    &\equiv
    \int \frac{d\hat{k}}{4\pi} \Big\{
    G_{\mu\nu;\lambda_1 \lambda_2}(\bphat_1, \bkhat) G_{\rho\sigma;\lambda_1 \lambda_2}(\bkhat, \bphat_3)\notag\\
    &\qquad\qquad \times \bar{u}_{\lambda_3}(p_3) \gamma^\rho \left(\gamma^0 + \hat{k}_j \gamma^j \right) \gamma^\mu u_{\lambda_1}(p_1)\notag\\
    &\qquad\qquad \times \bar{u}_{\lambda_4}(p_4) \gamma^{\sigma}\left( \gamma^0 - \hat{k}_l \gamma^l\right) \gamma^\nu u_{\lambda_2}(p_2)\Big\}\,.
    \label{eq:calT}
\end{align}

The one-loop beta function has been obtained in Refs.~\cite{Evans:1998ek,Schafer:1998na} (see Appendix~\ref{sec:IT} for the detailed derivation).
We consider the beta function for scattering of incoming quarks with the same helicities $\lambda_1 = \lambda_2 = 1/2$ ($++$) and with the opposite helicities $\lambda_1 = 1/2$, $\lambda_2 = -1/2$ ($+-$).
The result is expressed in terms of the partial wave expansion of the coupling function:
\begin{equation}
    G(\bphat, \bkhat) = \sum_L (2L+1) G_L P_L(\bphat \cdot \bkhat)\,,
\end{equation}
where $P_L(z)$ is the Legendre polynomial of a degree $L$.
We express the amplitude in terms of the coupling function in the orbital angular momentum channel $L$.

For temporal and spatial coupling function in the helicity $++$ channel, we obtain the RG equations:
\begin{align}
    \frac{d G^{0}_{L++} }{dt} &= -\frac{N}{4} \left[\left(G^{0}_{L++}\right)^2 +  2 G^0_{L++} G^i_{L++} + 5\left(G^i_{L++}\right)^2 \right]\,,
    \label{eq:RG0++}\\
    \frac{d G^i_{L++}}{dt} &= -\frac{N}{12} \left[
    \left(G^0_{L++}\right)^2 +10 G^0_{L++} G^i_{L++} + 13 \left(G^i_{L++}\right)^2 \right]\,,
    \label{eq:RGi++}
\end{align}
where $N=\mu^2 / (2\pi^2)$ is the density of states on the Fermi surface.
One can diagonalize these equations\footnote{Note the right hand side of the RG equations differ by a factor two from Refs.~\cite{Evans:1998ek, Evans:1998nf, Schafer:1998na} due to the difference in the convention of the color coefficient~\eqref{eq:cR}. In our convention, $c_R$ is twice as large compared to their convention.}:
\begin{align}
    \frac{d \left(G^0_{L++} + 3 G^i_{L++}\right) }{dt} &= -\frac{N}{2} \left(G^0_{L++} + 3 G^i_{L++} \right)^2\,, \label{eq:RGsinglet++}\\
    \frac{d \left(G^0_{L++} - G^i_{L++}\right)}{dt} &= -\frac{N}{6} \left(G^0_{L++} - G^i_{L++}\right)^2\,. \label{eq:RGtriplet++}
\end{align}
Likewise, the RG equations for the coupling in the $+-$ channel are
\begin{align}
    \frac{dG^0_{L+-}}{dt}
    &= -\frac{N}{4} \left[\left(G^0_{L+-}\right)^2 +  2 G^0_{L+-} G^i_{L+-} + \left(G^i_{L+-}\right)^2 \right]\,, \label{eq:RG0+-}\\
    \frac{dG^i_{L+-}}{dt} &= -\frac{N}{12} \left[
    \left(G^0_{L+-}\right)^2 + 2 G^0_{L+-} G^i_{L+-} + \left(G^i_{L+-}\right)^2 \right]\,.
    \label{eq:RGi+-}
\end{align}
These equations can be diagonalized as
\begin{align}
    \frac{d\left(G^0_{L+-}+ G^i_{L+-}\right)}{dt}
    &= -\frac{N}{3} \left(G^0_{L+-} + G^i_{L+-}\right)^2\,, \label{eq:RGsinglet+-}\\
    \frac{d\left(G^0_{L+-} - 3 G^i_{L+-}\right)}{dt} &= 0\,.
    \label{eq:RGtrivial+-}
\end{align}
The RG equations decouple among the different combinations of the coupling functions $G^0_{L++} + 3 G^i_{L++}$, $G^0_{L++} - G^i_{L++}$, $G^0_{L+} + G^i_{L+-}$, and $G^0_{L+-} - 3 G^i_{L+-}$.

Let us now clarify the meaning of these RG equations based on the tree-level amplitude.
The same classification was elaborated in the accompanying paper.
The tree-level amplitude~\eqref{eq:tree} in the helicity $++$ channel is
\begin{align}
    \calM_{++} (\theta)
    &= G^0_{++} \cos^2 \frac{\theta}2 + G^i_{++} \left(\cos^2 \frac{\theta}2 + 2 \sin^2 \frac{\theta}2\right) \,.
    \label{eq:M++}
\end{align}
and in the $+-$ channel is
\begin{align}
    \calM_{+-}(\theta) 
    &= G^0_{+-} \cos^2 \frac{\theta}2 + G^i_{+-} \cos^2 \frac{\theta}2  \,.
\end{align}

The projection onto the helicity eigenstate is done by the expansion in terms of the Wigner $d$-matrix~\cite{Jacob:1959at,Chung:1971ri}:
\begin{align}
    \mathcal{M}_{\lambda_1\lambda_2}(\theta) = \sum_J (2J+1)\mathcal{H}^J_{\lambda_1 \lambda_2} d^J_{\lambda \lambda'}(\theta) \,,
\end{align}
where $\lambda= \lambda_1 - \lambda_2$ and $\lambda' = \lambda_3 - \lambda_4$.
Now, we further decompose it in canonical partial-wave channels with orbital angular momentum $L$ and spin $S$.
\begin{align}
    |J;\lambda_1\lambda_2 \rangle = \sum_{LS} \sqrt{\frac{2L+1}{2J+1}} C_{L0S\lambda}^{J\lambda} C_{\frac12 \lambda_1 \frac12 -\lambda_2}^{S\lambda} |J;LS \rangle\,.
\end{align}
The helicity amplitudes in the $++$ channel are~\cite{Fujimoto:2025liq}
\begin{align}
    \mathcal{H}_{++}^{J,S=0,L=J} &= G^0_{L++} + 3G^i_{L++}\,,
    \label{eq:Hsinglet}\\
    \mathcal{H}_{++}^{J,S=1,L=J\pm 1} &= G^0_{L++} - G^i_{L++}\,.
    \label{eq:Htriplet}
\end{align}
As is evident from these expressions, the RG equation~\eqref{eq:RGsinglet++} corresponds to the spin singlet amplitude, while equation~\eqref{eq:RGtriplet++} corresponds to the spin triplet amplitude.

The helicity amplitude in the $+-$ channel with the total angular momentum $J$ is~\cite{Fujimoto:2025liq}
\begin{align}
    \mathcal{H}_{+-}^{J,S=1,L} &= G^0_{L+-} + G^i_{L+-}\,,
\end{align}
which is valid for $L=J-1$, $J$, and $J+1$.
This amplitude clearly corresponds to the RG equation~\eqref{eq:RGsinglet+-}.
The RG equation in the helicity $+-$ channel has non-trivial beta function only for $S=1$ component, and the other RG equation~\eqref{eq:RGtrivial+-} is trivial with a vanishing beta function.

\subsection{Self-energy correction}
\label{sec:self}

In the derivation in the previous subsection, we neglected HDL resummation in the quark propagator.
However, one has to resum a self-energy contribution $\Sigma$ in the quark propagator because both $k_0$ and $\epsilon_k$ in the propagator of the quasiparticle are equal or smaller than the soft scale $\sim g\mu$ in the problem: $i/\slashed{k} \to i/(\slashed{k} - \Sigma)$.
This quark self-energy correction amounts to the wave function renormalization.
Namely, the effect of including the self-energy is to replace
\begin{equation}
    k_0 \to \frac{k_0}{Z(k_0)}\,,
\end{equation}
in the quark propagator.
The quark wave function renormalization factor $Z(k_0)$ up to the leading logarithm is obtained as~\cite{Brown:1999yd, Manuel:2000nh,  Gerhold:2005uu}
\begin{equation}
    Z(k_0) \equiv \left(1 + \frac{g^2}{18\pi^2} \ln \frac{48 e^2 \mg^2}{\pi^2 k_0^2}\right)^{-1}\,,
\end{equation}
where $\mg^2= \mD^2 / 3$ is the effective gluon mass at finite density, and $e$ is the base of the natural logarithm.

\subsubsection{Resummed quark propagator}

Now, we include the wave function renormalization effect up to $O(g^2)$ in the RG equation.
This effect only appears through the integral $\mathcal{I}$ defined in Eq.~\eqref{eq:calI}.
We replace $\mathcal{I}$ with $\mathcal{I}'$:
\begin{align}
    \mathcal{I}' &=-\frac14  \int_{d\Lambda} \frac{k^2 \, d k}{2\pi^2}\int_{-\infty}^{\infty} \frac{dk_0}{2\pi}\,
    \frac{1}{\left[k_0/Z(k_0)\right]^2 - \epsilon_k^2}\,.
\end{align}
We evaluate the $k_0$-integral by performing the Wick rotation $k_0 \to i k_4$:
\begin{align}
    i\int_{-\infty}^{\infty} \frac{dk_4}{2\pi}\, \frac{Z^2(ik_4)}{\left(i k_4 \right)^2+ \left[Z(ik_4) \epsilon_k\right]^2}
    = -\frac{i Z(\epsilon_k)}{2 |\epsilon_k|}\,.    
\end{align}
Then, the integral $\mathcal{I}'$ is
\begin{align}
    \mathcal{I}'
    &= -\frac{N}{4} \int_{d\Lambda} d\epsilon_k\, \left[-\frac{i Z(\epsilon_k)}{2 |\epsilon_k|}  \right] \,,\notag\\
    &= \frac{i N}{4} \frac{9 \pi^2}{g^2} \ln \left[\frac{1 + \frac{g^2}{18\pi^2} \ln \left(\frac{48 e^2 \mg^2}{\pi^2 \Lambda^2}\right) +\frac{g^2}{9\pi^2}t}{1 + \frac{g^2}{18\pi^2} \ln \left(\frac{48 e^2 \mg^2}{\pi^2 \Lambda^2}\right)}\right]\,.
\end{align}
Taking the derivative with respect to $t$, one gets
\begin{align}
    \frac{d \mathcal{I}'}{d t} = \frac{i N}{4} \frac{1}{1 + \frac{g^2}{9\pi^2} t + O(g^2)}\,.
    \label{eq:calIprime}
\end{align}
We see below that the $t$-independent term of $O(g^2)$ gives a higher-order contribution to the gap compared to the $t$-dependent term of $O(g^2)$.

In the beta function, the density of states is replaced by
\begin{align}
    N \to Z_{\psi}(t) N\,,\quad
    Z_{\psi}(t) \equiv \left(1 + \frac{g^2}{9 \pi^2}t \right)^{-1}\,.
\end{align}
Although the term proportional to $t$ contributes at $O(g^2)$, this is important as the $t$-dependence changes the solution of the RG equation entirely.
We dropped the constant term at $O(g^2)$ as these corrections remain to be small.

\subsubsection{Role of quark self energy in the gap}

Let us now illustrate the role of the quark self energy in the pairing gap by the following simple argument.

The gap is given by the singularity in the solution of the RG equation.
Let us consider the simple RG equation without the tree-level beta function.
\begin{equation}
    \frac{d G}{dt} = -N G^2\,.
\end{equation}
Its solution at general $t>0$ is
\begin{align}
    G(t) = \frac{G(0)}{1 + N G(0) t}\,.
\end{align}
When $G(0) < 0$, this solution has a singularity at
\begin{align}
    t = \frac{1}{N|G(0)|}\,.
\end{align}
The energy scale $\Lambda$ at the singularity can be regarded as the pairing gap.
From the relation $t = - \ln(\Lambda / \mD)$, one finds
\begin{align}
    \ln \left(\frac{\Delta}{\mD} \right) = -\frac{1}{N|G(0)|}\,.
\end{align}

Now we do the same for the RG equation including the self-energy corrections:
\begin{align}
    \frac{d G}{dt} = - \frac{N}{1 + c\epsilon + \epsilon t} G^2\,,
\end{align}
where $\epsilon \ll 1$ is a small parameter related to the wave function renormalization.
The term $c\epsilon$ in the denominator corresponds to the $t$-independent term of $O(g^2)$ in the denominator of Eq.~\eqref{eq:calIprime}.
Its solution is
\begin{align}
    G(t) = \frac{G(0)}{1 + \frac{N G(0)}{\epsilon} \ln\left(\frac{1 + c\epsilon + \epsilon t}{1 + c \epsilon}\right)}\,.
\end{align}
When $G(0)<0$, this solution is singular at
\begin{align}
    0&=1 + \frac{N G(0)}{\epsilon} \ln\left(\frac{1 + c\epsilon + \epsilon t}{1 + c \epsilon}\right)\notag\\
    &\simeq 1 - N |G(0)| \left[(1 - c\epsilon)t - \frac12\epsilon t^2 \right]
\end{align}
One can solve
\begin{align}
    t
    &\simeq \frac{1}{N|G(0)|} + \frac{c\epsilon}{N|G(0)|} + \frac{\epsilon}{2N^2|G(0)|^2} +O(\epsilon^2)\,.
\end{align}
The gap is
\begin{align}
    \ln \left(\frac{\Delta}{\mD} \right) = -\frac{1}{N|G(0)|} - \frac{c\epsilon}{N|G(0)|} - \frac{\epsilon}{2N^2|G(0)|^2}\,.
\end{align}
Now, there is an additional suppression factor arising from the wave function renormalization appearing in the gap, which corresponds the second and third terms in the above expression.
As we see below, in the case of QCD with a magnetic interaction, the variables scale as $|G(0)| \sim g$ and $\epsilon \sim g^2$ so that the leading term scales as $1/(N|G(0)|) \sim O(1/g)$.
The subleading corrections scale as $c\epsilon / (N|G(0)|) \sim O(g)$ and $\epsilon/(2N^2 |G(0)|^2) \sim O(g^0)$.
The term dependent on $c$, which appears on the second term on the right-hand side, is one order higher in $g$ compared to the third term.
Therefore, second term is subleading, and can be neglected if we only keep up to the $O(g^0)$ terms in the gap calculation, which is the reliable order in the current setup.
This justifies that we only keep the $t$-dependent term in the denominator in Eq.~\eqref{eq:calIprime}.

\subsection{Renormalization of the tree-level amplitude}
\label{sec:tree}

Now, we illustrate how tree-level amplitude for the unscreened magnetic field gives the nonzero beta function in the RG transformation.
In the context of dense QCD, this was first pointed out by Son in Ref.~\cite{Son:1998uk}.
Although we do not delve into the topic, the very existence of this term, which also survives in the forward scattering limit with $\bphat_1 \cdot \bphat_2 = \bphat_3 \cdot \bphat_4$, is responsible for the non-Fermi liquid nature of quark matter~\cite{Schafer:2004zf}.

\subsubsection{Statically screened electric interaction}

The tree-level beta function vanishes for the interaction that is screened statically.
From the chromoelectric gluon self-energy~\eqref{eq:PiE}, the tree-level amplitude in the electric sector is
\begin{align}
    G^0(q) = -\frac{c_R g^2}{q_0^2 - q^2 - \mD^2 + i0^+}\,.
\end{align}
When one limits the mode near the Fermi surface $|\epsilon_p| < \Lambda$, then the energy transfer can only be limited to $|q_0| < \Lambda$.
That means, the tree-level amplitude at a given scale $\Lambda$ is
\begin{align}
    G^0(\theta; \Lambda) \simeq -\frac{c_R g^2}{\Lambda^2 - 2\mu^2(1-\cos\theta) - \mD^2 + i0^+}\,.
\end{align}
We expand the coupling function in the partial wave channel with angular momentum $L$,
\begin{align}
    G^0_L(\Lambda) = \frac12 \int_{0}^\pi d\theta\,\sin\theta\, P_L(\cos\theta) G^0(\theta; \Lambda)\,.
\end{align}
When performing the RG transformation $\Lambda \to e^{-t} \Lambda$, then the rate of change in the tree-level amplitude in the partial wave channel can be evaluated as
\begin{align}
    \beta^0_{\mathrm{tree}} = \frac{d G^0_L(e^{-t} \Lambda)}{d t}
    = 0 + O \left(\frac{e^{-2t}\Lambda^2}{\mD^2}\right)\,.
\end{align}
There is only a contribution at $O(e^{-2t})$, which is only an irrelevant piece in the beta function.

\subsubsection{Unscreened magnetic interaction}

The tree level amplitude of unscreened magnetic sector is
\begin{align}
    G^i(q) = -\frac{c_R g^2}{q_0^2 - q^2 + i0^+}\,.
\end{align}
Similar to the electric case, the tree-level amplitude at a given scale $\Lambda$ is
\begin{align}
    G^i(\theta; \Lambda) \simeq  -\frac{c_R g^2}{\Lambda^2 - 2\mu^2(1-\cos\theta) + i0^+}\,.
\end{align}
In the partial wave channel with angular momentum $L$,
\begin{align}
    G^i_L(\Lambda) = \frac12 \int_{0}^\pi d\theta\,\sin\theta\, P_L(\cos\theta) G^i(\theta; \Lambda)\,,
    \label{eq:partial}
\end{align}
the rate of change in the tree-level amplitude when performing the RG transformation $\Lambda \to e^{-t} \Lambda$ is 
\begin{align}
    \beta^i_{\mathrm{tree}} = \frac{d G^i_L(e^{-t} \Lambda)}{d t}
    = \frac{c_R g^2}{2\mu^2} + O\left(\frac{e^{-2t} \Lambda^2}{\mu^2}\right)\,.
\end{align}
We have suppressed the irrelevant piece in the beta function.
This beta function is dominated by the contribution from the integral boundary $\theta \simeq 0$ in Eq~\eqref{eq:partial}.
Note that the marginal piece of the tree-level beta function does not depend on the angular momentum $L$.

\subsubsection{Dynamical screening from Landau damping}

In reality, magnetic interaction is also screened dynamically, depending on $q_0$, due to the Landau damping.
Therefore, the tree-level beta function, which is the rate of change in the magnetic interaction, of the dynamically screened interaction becomes smaller than that of the unscreened interaction.
When evaluating the tree-level beta function, incorporating the correct frequency dependence of the amplitude is important.
As we saw above, for $q_0 \ll |\bq|$ and to leading order in perturbation theory, the gluon self-energy in the magnetic sector $\Pi^{\mathrm{M}}$ is given by Eq.~\eqref{eq:PiM}, which is proportional to $\sim q_0$.
Therefore, this term is dominant compared to the term $q_0^2$ in the denominator of $G^i$.
Now, the tree-level amplitude of the magnetic sector becomes
\begin{align}
    G^i(q) = -\frac{c_R g^2}{q_0^2 - q^2 +i \frac{\pi}4\mD^2\frac{q_0}{q} + i0^+}\,.
\end{align}
This equation shows that there is a dynamical screening with a frequency-dependent cutoff $q_c \simeq (\pi \mD^2 q_0 /4)^{1/3}$.
At a given renormalization scale $\Lambda$, the tree-level amplitude is
\begin{align}
    G^i(\theta; \Lambda) \simeq \frac{c_R g^2}{2\mu^2(1-\cos\theta) - i\frac{\pi}4\mD^2\frac{\Lambda}{\sqrt{2\mu(1-\cos\theta)}} - i0^+}\,,
\end{align}
where we only kept the dominant $\Lambda$ contribution in the denominator.
Because $\Lambda \simeq \mD$, the cutoff is at $q_c \simeq (\pi /4)^{1/3} \mD$.
In the partial wave channel, the tree-level beta function in the partial wave channel $L$ is
\begin{align}
    \beta^i_{\mathrm{tree}} = \frac{d G^i_L(e^{-t} \Lambda)}{d t}
    = \frac{c_R g^2}{6\mu^2} + O\left(\frac{e^{-2t/3} \Lambda^{2/3}}{\mu^{2/3}}\right)\,.
    \label{eq:betatree}
\end{align}
Again, the marginal part of the tree-level beta function does not depend on the angular momentum $L$.

\subsection{Full RG equation}
\label{sec:fullRG}

By combining the results obtained in this section, the full RG equation for the linear combination of the coupling functions, $G^0 + \alpha G^i$, can be schematically summarized as
\begin{align}
    \frac{d \left(G^0_{L\lambda_1\lambda_2} +\alpha G^i_{L\lambda_1\lambda_2}\right)}{dt}
    = Z_\psi(t) \delta\beta\left[G^0_{L\lambda_1\lambda_2} + \alpha G^i_{L\lambda_1\lambda_2}\right] + \alpha \beta^i_{\mathrm{tree}}\,,
    \label{eq:RGschematic}
\end{align}
where we obtained the one-loop beta function $\delta \beta$ as a functional of $G^0(t)$ and $G^i(t)$ in Sec.~\ref{sec:oneloop}, the self-energy correction $Z_\psi(t)$ in Sec.~\ref{sec:self}, and the tree-level beta function $\beta^i_{\mathrm{tree}}$ in Sec.~\ref{sec:tree}.

For the initial helicities $(\lambda_1, \lambda_2) = (+,+)$, the concrete expression of the RG equations is
\begin{align}
    \frac{d \left(G^0_{L++} + 3 G^i_{L++}\right) }{dt} &= -\frac{N\left(G^0_{L++} + 3 G^i_{L++} \right)^2}{2\left(1 + \frac{g^2}{9 \pi^2}t \right)}  + \frac{c_R g^2}{2\mu^2}\,,\\
    \frac{d \left(G^0_{L++} - G^i_{L++}\right)}{dt} &= -\frac{N\left(G^0_{L++} - G^i_{L++}\right)^2}{6\left(1 + \frac{g^2}{9 \pi^2}t \right)}  - \frac{c_R g^2}{6\mu^2}\,,
\end{align}
and for the initial helicities $(\lambda_1, \lambda_2) = (+,-)$, the RG equations are
\begin{align}
    \frac{d\left(G^0_{L+-}+ G^i_{L+-}\right)}{dt}
    &= -\frac{N\left(G^0_{L+-}+ G^i_{L+-}\right)^2}{3\left(1 + \frac{g^2}{9 \pi^2}t \right)}  + \frac{c_R g^2}{6\mu^2}\,, \\
    \frac{d\left(G^0_{L+-} - 3 G^i_{L+-}\right)}{dt} &= - \frac{c_R g^2}{2\mu^2}\,.
\end{align}
The last equation does not lead to BCS instability in the solution.
This corresponds to the fact that the spin singlet state is missing in the helicity $+-$ amplitude.

\section{Color-superconducting gap from the RG equation}
\label{sec:solution}

In this section, we solve the RG equation as an initial value problem and find the pole in the solution; we identify such a pole as the energy scale at which BCS instability kicks in.
We impose the initial condition at $t=0$, which is set by the matching condition to perturbation theory~\eqref{eq:matching} at a soft scale $\sim \mD$, and solve for  arbitrary $t = -\ln(\Lambda / \mD) > 0$.
The solution at arbitrary $t$ hits the singularity at a certain $t = t_{\mathrm{BCS}}$, which is the manifestation of BCS instability in the Landau Fermi liquid description of the dense matter.
The superconducting gap is the energy scale at which the Fermi liquid description breaks down, i.e., $\Delta \simeq \mD e^{-t_{\mathrm{BCS}}}$.
We obtain the perturbative series for logarithm of the pairing gap up to the next-to-leading order $O(g^0)$.
We show the expressions for each color, flavor, and angular momentum channels as classified in Table~\ref{tab:gap}.

So far, we have discussed the RG equations in Minkowski space, but the RG equations remain the same in Euclidean space because the RG equation is derived in Sec.~\ref{sec:Fermi} from the consistency of the partition function, which remains the same after the Wick rotation.
Hereafter, we will work in Euclidean space by analytically continuing $p_0 \to i p_4$ so that the calculation related with the dynamical screening becomes cleaner.

For the sake of simplicity, let us redefine the coupling function as
\begin{align}
    G_L \to V_L \equiv N G_L\,,
\end{align}
by absorbing the density of states $N$ into the definition of the coupling function.
Then, the RG equations become
\begin{align}
    \frac{d \Vsing_{L++} }{dt} &= -\frac{\left(\Vsing_{L++} \right)^2}{2\left(1 + \frac{g^2}{9 \pi^2}t \right)}  + \frac{c_R g^2}{4\pi^2}\,, \label{eq:RGVs++}\\
    \frac{d \Vtrip_{L++}}{dt} &= -\frac{\left(\Vtrip_{L++}\right)^2}{6\left(1 + \frac{g^2}{9 \pi^2}t \right)}  - \frac{c_R g^2}{12\pi^2}\,, \label{eq:RGVt++}
\end{align}
where $\Vsing_{L++} \equiv V^0_{L++} + 3V^i_{L++}$ and $\Vtrip_{L++} \equiv V^0_{L++} - V^i_{L++}$ are the spin singlet and triplet combination of coupling functions, respectively.
And for the initial helicities $(\lambda_1, \lambda_2) = (+,-)$, the RG equations are
\begin{align}
    \frac{d\Vtrip_{L+-}}{dt}
    &= -\frac{\left(\Vtrip_{L+-}\right)^2}{3\left(1 + \frac{g^2}{9 \pi^2}t \right)}  + \frac{c_R g^2}{12\pi^2}\,, \label{eq:RGVt+-}
\end{align}
where $\Vtrip_{L+-} \equiv V^0_{L+-} + V^i_{L+-}$ is the spin triplet coupling function.

\subsection{Initial condition}

The initial condition at $t=0$ is set by the matching condition~\eqref{eq:matching}.
In this section, we discuss how to obtain the initial condition in a given partial wave $L$.

The initial condition for the electric coupling function is
\begin{align}
    V^0_L(t=0)
    = \frac{c_R g^2}{8\pi^2} \int_{0}^{\pi} d\theta \, \sin\theta\, P_L(\cos\theta) \frac{1}{1-\cos\theta + 2\OmegaE^2}\,,
\end{align}
where we define $\OmegaE^2 \equiv \mD^2 / (4\mu^2) = N_f g^2 / (8\pi^2)$.
For the magnetic amplitude, first one has to find the cutoff for the angular integration.
The magnetic propagator reads
\begin{align}
    \DM(iq_4, q) \simeq \frac{1}{q^2 + \frac{\pi}{4} \mD^2 \frac{q_4}{q}}\,.
\end{align}
Because in the effective theory near the Fermi surface, the energy transfer $q_4$ is limited to the scale $q_4 \lesssim \Lambda$, and the scale parameter $t$ quantifies the departure from the soft scale, i.e. $t = -\ln(\Lambda /\mD)$, in the above equation $q_4 \simeq \mD$ has to be substituted for the initial condition at $t=0$.
Then, the initial condition for the magnetic coupling function is
\begin{align}
    V^i_L(t=0)
    \simeq \frac{c_R g^2}{8\pi^2} \int_{0}^{\pi} d\theta \, \sin\theta\, P_L(\cos\theta) \frac{1}{1-\cos\theta + \frac{\pi \OmegaE^3}{\sqrt{2 (1-\cos\theta)}} } \,.
\end{align}

For $L=0$ channel to leading order, one finds
\begin{align}
    V_{L=0}^0(t=0) &= \frac{c_R g^2}{8\pi^2} \ln\left(1 + \OmegaE^{-2}\right)\,,\\
    V_{L=0}^i(t=0) &= \frac{c_R g^2}{12\pi^2} \ln\left(1 + \frac{4}{\pi} \OmegaE^{-3}\right) \,.
\end{align}
For $L=1$ channel to leading order,
\begin{align}
    V_{L=1}^0(t=0) &= \frac{c_R g^2}{8\pi^2} \left[-2 + \ln\left(1 + \OmegaE^{-2}\right)\right] + O(g^4)\,,\\
    V_{L=1}^i(t=0) &= \frac{c_R g^2}{12\pi^2} \left[-3 + \ln\left(1 + \frac{4}{\pi} \OmegaE^{-3}\right)\right] + O(g^4)\,.
\end{align}
The initial conditions at $t=0$ for the RG equations for the spin singlet and helicity $++$ channel are
\begin{align}
    \Vsing_{0++}(t=0) &= \frac{c_R g^2}{4\pi^2} \ln\left[\left(1 + \OmegaE^{-2}\right)^{\frac12} \left(1 + \frac{4}{\pi} \OmegaE^{-3}\right) \right]\,, \label{eq:Vs0++}\\
    \Vsing_{1++}(t=0) &= \frac{c_R g^2}{4\pi^2} \ln\left[e^{-4}\left(1 + \OmegaE^{-2}\right)^{\frac12} \left(1 + \frac{4}{\pi} \OmegaE^{-3}\right) \right]\,, \label{eq:Vs1++}
\end{align}
where the subscript $0$ and $1$ refers to the orbital angular momentum $L=0$ and $1$.
For the spin triplet and helicity $+-$ channel, they are
\begin{align}
    \Vtrip_{0++}(t=0) &= -\frac{c_R g^2}{12\pi^2} \ln\left[\left(1 + \OmegaE^{-2}\right)^{-\frac32} \left(1 + \frac{4}{\pi} \OmegaE^{-3}\right) \right]\,, \label{eq:Vt0++}\\
    \Vtrip_{1++}(t=0) &= \Vtrip_{0++}(t=0)\,. \label{eq:Vt1++}
\end{align}
Note that the sign of the coefficient is negative.
This means that the amplitude is attractive for color $\boldsymbol{6}$ channel with $c_R = 1/3$.
This was first noticed in Ref.~\cite{Schafer:2006ue}.
As we will see below, much stronger attraction is powered by the unscreened magnetic interaction.
Finally, for the helicity $+-$ channel, there is only spin-triplet channel:
\begin{align}
    \Vtrip_{0+-}(t=0) &= \frac{c_R g^2}{12\pi^2} \ln\left[\left(1 + \OmegaE^{-2}\right)^{\frac32} \left(1 + \frac{4}{\pi} \OmegaE^{-3}\right) \right]\,. \label{eq:Vt0+-}
\end{align}
When $g$ is small enough, the factor $1$ in the above expressions may be neglected, i.e., $1 + \OmegaE^{-2} \simeq \OmegaE^{-2}$.
Nevertheless, we will keep this factor $1$ for completeness.

\subsection{General solution of the RG equation}

We solve the following differential equation, which has the identical structure as in Eqs.~(\ref{eq:RGVs++}--\ref{eq:RGVt+-}), under the initial condition $y(t=0) = - b g^2 l / \pi^2$:
\begin{equation}
    \frac{d y(t)}{dt} = - \frac{a}{1 + \frac{g^2}{9\pi^2} t} y^2(t) - \frac{b g^2}{\pi^2}\,.
\end{equation}
We rescale $\gbar = g / (18\pi\sqrt{ab})$ so that the equation now becomes
\begin{equation}
    \frac{d y(t)}{dt} = - \frac{a}{h^2(t)} y^2(t) - \frac{c^2}{4a} \gbar^2\,,
    \label{eq:generalRG}
\end{equation}
where we define the constant $c = 36ab$ and the function $h(t) = \sqrt{1 + c \gbar^2 t}$.
The solution is
\begin{equation}
    y(t) = -\frac{c \gbar h(t) \mathcal{X}}{2a \mathcal{Y}}\,,
    \label{eq:generalsol}
\end{equation}
where the numerator $\mathcal{X}$ and denominator $\mathcal{Y}$ are
\begin{align}
    \label{eq:calX}
    \mathcal{X}
    &= \left[c\gbar l J_0\left(\frac{1}{\gbar}\right) - 2 J_1\left(\frac{1}{\gbar}\right) \right] Y_1\left(\frac{h(t)}{\gbar}\right)
    - \left[c\gbar l Y_0\left(\frac{1}{\gbar}\right) - 2 Y_1\left(\frac{1}{\gbar}\right) \right] J_1\left(\frac{h(t)}{\gbar}\right)\,,\\
    \mathcal{Y}
    &= \left[c\gbar l J_0\left(\frac{1}{\gbar}\right) - 2 J_1\left(\frac{1}{\gbar}\right) \right] Y_0\left(\frac{h(t)}{\gbar}\right)
    - \left[c\gbar l Y_0\left(\frac{1}{\gbar}\right) - 2 Y_1\left(\frac{1}{\gbar}\right) \right] J_0\left(\frac{h(t)}{\gbar}\right)\,.
    \label{eq:calY}
\end{align}
The Fermi liquid description breaks down when $\mathcal{Y} = 0$, and we denote such $t$ as $t_{\mathrm{BCS}}$.
We expand $\mathcal{Y}$ by using asymptotic expansions for the Bessel function for large argument $x \gg 1$ up to $O(1/x)$:
\begin{align}
    J_\nu(x) &\sim \sqrt{\frac{2}{\pi x}} \bigg[\cos\left(x - \frac{\nu\pi}{2} - \frac{\pi}{4}\right)
    - \frac{4\nu^2 - 1}{8x} \sin\left(x - \frac{\nu\pi}{2} - \frac{\pi}{4}\right) \bigg]\,, \label{eq:BesselJ}\\
    Y_\nu(x) &\sim \sqrt{\frac{2}{\pi x}} \bigg[\sin\left(x - \frac{\nu\pi}{2} - \frac{\pi}{4}\right)
    - \frac{4\nu^2 - 1}{8x} \cos\left(x - \frac{\nu\pi}{2} - \frac{\pi}{4}\right)\bigg]\,.\label{eq:BesselY}
\end{align}
Substituting these into the equation $\mathcal{Y}=0$, one obtains
\begin{align}
    \tan\left(\frac{h(t_{\mathrm{BCS}}) - 1}{\gbar}\right) \simeq \frac{2}{(-1 + c l) \gbar} + O(\gbar)\,.
    \label{eq:tanBCS}
\end{align}
Details of the computation is in Appendix~\ref{sec:solRGeq}.
By taking $\arctan$ in the both-hand sides, we obtain
\begin{align}
    \frac{\sqrt{1 + c\gbar^2 t_{\mathrm{BCS}}} - 1}{\gbar}
    &\simeq \frac{\pi}{2} + \frac12 (1 - cl) \gbar\,,
\end{align}
where we used the relation $\arctan(1/x) = \pi/2 - \arctan(x)$ and expanded as $\arctan(x) \simeq x$.
The singularity, which corresponds to BCS instability, is found at
\begin{align}
    t_{\mathrm{BCS}}
    &\simeq \frac{\pi}{c \gbar} + \frac{1}{c}\left(1 + \frac{\pi^2}{4}\right) - l\,, \notag \\
    &\simeq \frac{\pi^2}{2\sqrt{ab}g} + \frac{\pi^2 + 4}{144ab} - l\,.
\end{align}
The magnitude of the first term is determined by the angular momentum structure, which is captured by the factor $a$, and the strength of the magnetic interaction, captured by the factor $b$.
The second term arises from the self-energy correction;
the factor $\pi^2 + 4$ is common for the solution of the gap equation~\cite{Wang:2001aq}.
The third term arises from the initial condition for the solution of the RG equation.
From the relation $\ln(\Delta / \mD) \simeq -t_{\mathrm{BCS}}$, the series expansion of the logarithm of the gap is
\begin{align}
    \ln\left(\frac{\Delta}{\mu}\right) = - \frac{\pi^2}{2\sqrt{ab}g} - \frac{\pi^2 + 4}{144ab} + l + \ln(2\OmegaE)\,.
    \label{eq:lngap}
\end{align}
This series expansion can be reliable up to $O(g^0)$.
Beyond this order, there are other sources of the perturbative corrections that are not currently taken into account.

\subsection{The gap in each color, flavor, helicity, and angular momentum channel}
\label{sec:classify}

In this section, let us consider the two-flavor case.
As each quark is a flavor doublet $\boldsymbol{2}$, the pair can form $\boldsymbol{2} \otimes \boldsymbol{2} = \boldsymbol{1} \oplus \boldsymbol{3}$: the flavor singlet $\boldsymbol{1}_f$ is an antisymmetric, and $\boldsymbol{3}_f$ is a symmetric representation.

\begin{table}[h]
    \centering
    \begin{tabular}{c|ccccc}
        Case & Color & Flavor & Helicity & $^{2S+1}L_J$ & Gap \\ \hline\hline
        I   & $\boldsymbol{\bar{3}}_c$ & $\boldsymbol{1}_f$ & $++$ & $^1S_0$ & \eqref{eq:gap1S0}\\
        II  & $\boldsymbol{\bar{3}}_c$ & $\boldsymbol{3}_f$ & $++$ & $^1P_1$ & \eqref{eq:gap1P1} \\
        III & $\boldsymbol{\bar{3}}_c$ & $\boldsymbol{3}_f$ & $+-$ & $^3S_1$ & \eqref{eq:gap3S1+-} \\
        IV  & $\boldsymbol{6}_c$       & $\boldsymbol{3}_f$ & $++$ & $^3P_0$ & \eqref{eq:gap3P0} \\
        V   & $\boldsymbol{6}_c$       & $\boldsymbol{1}_f$ & $++$ & $^3S_1$ & \eqref{eq:gap3S1++}
    \end{tabular}
    \caption{The possible attractive channel for a given color and flavor representation.}
    \label{tab:gap}
\end{table}

From the classification of the helicity amplitude in the accompanying paper~\cite{Fujimoto:2025liq}, candidates of the most attractive channel for each color-flavor channel are listed in Table~\ref{tab:gap}.
In the case of color-flavor $(\boldsymbol{\bar{3}}_c, \boldsymbol{3}_f)$ channel, there are two possible channel that can form the diquark condensate.
From the consideration of the helicity amplitude alone, one cannot determine which diquark channel is favored, and that should be determined by comparing the magnitude of the gap.

\begin{figure}
    \centering
    \includegraphics[width=0.5\columnwidth]{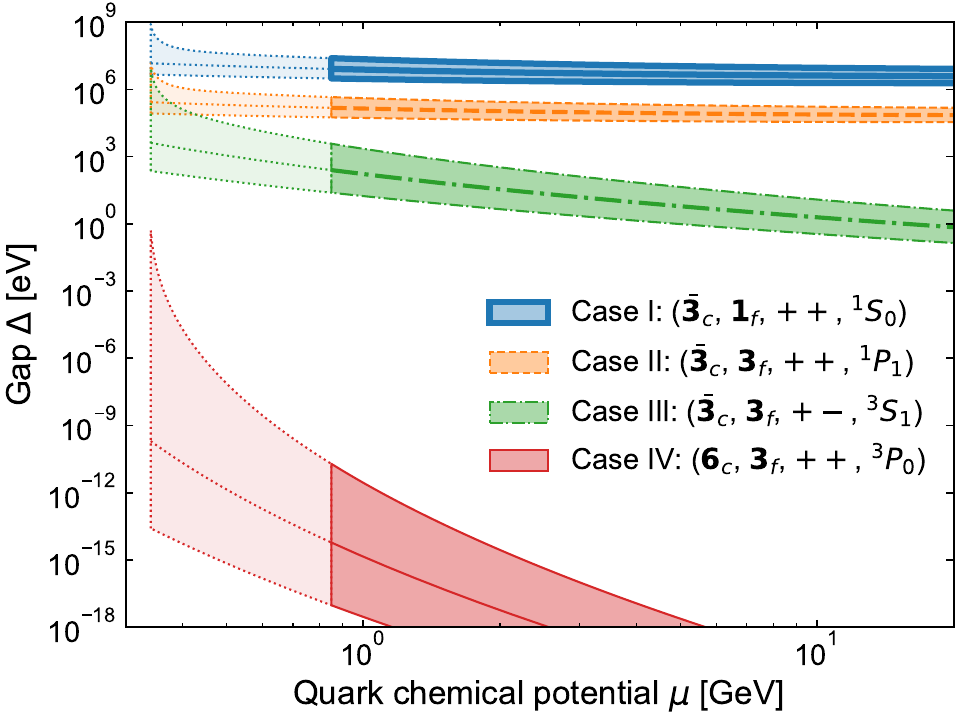}
    \caption{The comparison of the gap in different channels. Each channel is labeled by (color representation, flavor representation, helicity of incoming quarks, term symbol ${}^{2S+1}L_J$) as tabulated in Table~\ref{tab:gap}.}
    \label{fig:gap_compare}
\end{figure}

In Fig.~\ref{fig:gap_compare}, we compare the gap in different color, flavor, helicity and angular momentum channels; in the figure, each channel is labeled by these four parameters.
The explicit form of the one-loop results are listed below.
For the coupling constant $g$, we use the running coupling from the two-loop QCD beta function evaluated in the $\overline{\mathrm{MS}}$ scheme with the scale $\Lambda_{\overline{\mathrm{MS}}} \simeq 340\,\text{MeV}$.
The band corresponds to the scale variation, for which we follow the convention by choosing the renormalization scale $\bar{\Lambda} = 2\mu$ and varying it by a factor two.

\subsubsection*{Case I: $(\boldsymbol{\bar{3}}_c, \boldsymbol{1}_f, ++, {}^1S_0)$}

In this case, the corresponding RG equation is Eq.~\eqref{eq:RGVs++}, and the initial condition is set by Eq.~\eqref{eq:Vs0++}.
Then, the parameters in Eq.~\eqref{eq:lngap} are
\begin{align*}
    a = \frac12\,,\quad
    b = \frac16\,,\quad
    l = \ln\left[\left(1 + \OmegaE^{-2}\right)^{\frac12} \left(1 + \frac{4}{\pi} \OmegaE^{-3}\right) \right]\,.
\end{align*}
Then, the perturbative series for the gap parameter up to $O(g^0)$ is
\begin{align}
    \ln\left(\frac{\Delta}{\mu}\right) = - \frac{\sqrt{3}\pi^2}{g} - \frac{\pi^2 + 4}{12} + \ln\left[2\OmegaE\left(1 + \OmegaE^{-2}\right)^{\frac12} \left(1 + \frac{4}{\pi} \OmegaE^{-3}\right) \right]\,.
    \label{eq:gap1S0}
\end{align}

\subsubsection*{Case II: $(\boldsymbol{\bar{3}}_c, \boldsymbol{3}_f, ++, {}^1P_1)$}

In this case, the corresponding RG equation is Eq.~\eqref{eq:RGVs++}, and the initial condition is set by Eq.~\eqref{eq:Vs1++}.
Then, the parameters in Eq.~\eqref{eq:lngap} are
\begin{align*}
    a = \frac12\,,\quad
    b = \frac16\,,\quad
    l = \ln\left[e^{-4}\left(1 + \OmegaE^{-2}\right)^{\frac12} \left(1 + \frac{4}{\pi} \OmegaE^{-3}\right) \right]\,.
\end{align*}
Then, the perturbative series for the gap parameter up to $O(g^0)$ is
\begin{align}
    \ln\left(\frac{\Delta}{\mu}\right) = - \frac{\sqrt{3}\pi^2}{g} - \frac{\pi^2 + 4}{12} + \ln\left[2e^{-4} \OmegaE\left(1 + \OmegaE^{-2}\right)^{\frac12} \left(1 + \frac{4}{\pi} \OmegaE^{-3}\right) \right]\,.
    \label{eq:gap1P1}
\end{align}

\subsubsection*{Case III: $(\boldsymbol{\bar{3}}_c, \boldsymbol{3}_f, +-, {}^3S_1)$}

In this case, the corresponding RG equation is Eq.~\eqref{eq:RGVt+-}, and the initial condition is set by Eq.~\eqref{eq:Vt0++}.
Then, the parameters in Eq.~\eqref{eq:lngap} are
\begin{align*}
    a = \frac13\,,\quad
    b = \frac1{18}\,,\quad
    l = \ln\left[\left(1 + \OmegaE^{-2}\right)^{\frac32} \left(1 + \frac{4}{\pi} \OmegaE^{-3}\right) \right]\,.
\end{align*}
Then, the perturbative series for the gap parameter up to $O(g^0)$ is
\begin{align}
    \ln\left(\frac{\Delta}{\mu}\right) = - \frac{3\sqrt{3}\pi^2}{\sqrt{2} g} - \frac{3(\pi^2 + 4)}{8} + \ln\left[2\OmegaE \left(1 + \OmegaE^{-2}\right)^{\frac32} \left(1 + \frac{4}{\pi} \OmegaE^{-3}\right) \right]\,.
    \label{eq:gap3S1+-}
\end{align}
This gap in the case III is smaller than that of the case II.
This is evident from that the magnitude of the leading exponent is larger than the one above, i.e., $3\sqrt{3}\pi^2 / (\sqrt{2}g) > \sqrt{3}\pi^2 / g$, so the gap in the case III is exponentially suppressed.
One can also see this in Fig.~\ref{fig:gap_compare}.
Therefore, the case II is favored over the case III.

\subsubsection*{Case IV: $(\boldsymbol{6}_c, \boldsymbol{3}_f, ++, {}^3P_0)$}

In this case, the corresponding RG equation is Eq.~\eqref{eq:RGVt++}, and the initial condition is set by Eq.~\eqref{eq:Vt1++}.
Then, the parameters in Eq.~\eqref{eq:lngap} are
\begin{align*}
    a = \frac16\,,\quad
    b = \frac1{36}\,,\quad
    l = \ln\left[\left(1 + \OmegaE^{-2}\right)^{-\frac32} \left(1 + \frac{4}{\pi} \OmegaE^{-3}\right) \right]\,.
\end{align*}
Then, the perturbative series for the gap parameter up to $O(g^0)$ is
\begin{align}
    \ln\left(\frac{\Delta}{\mu}\right) = - \frac{3\sqrt{6}\pi^2}{g} - \frac{3(\pi^2 + 4)}{2} + \ln\left[2\OmegaE \left(1 + \OmegaE^{-2}\right)^{-\frac32} \left(1 + \frac{4}{\pi} \OmegaE^{-3}\right) \right]\,.
    \label{eq:gap3P0}
\end{align}

\subsubsection*{Case V: $(\boldsymbol{6}_c, \boldsymbol{1}_f, ++, {}^3S_1)$}

In this case, the corresponding RG equation is Eq.~\eqref{eq:RGVt++}, and the initial condition is set by Eq.~\eqref{eq:Vt0++}.
Then, the parameters in Eq.~\eqref{eq:lngap} are
\begin{align*}
    a = \frac16\,,\quad
    b = \frac1{36}\,,\quad
    l = \ln\left[\left(1 + \OmegaE^{-2}\right)^{-\frac32} \left(1 + \frac{4}{\pi} \OmegaE^{-3}\right) \right]\,.
\end{align*}
Then, the perturbative series for the gap parameter up to $O(g^0)$ is
\begin{align}
    \ln\left(\frac{\Delta}{\mu}\right) = - \frac{3\sqrt{6}\pi^2}{g} - \frac{3(\pi^2 + 4)}{2} + \ln\left[2\OmegaE \left(1 + \OmegaE^{-2}\right)^{-\frac32} \left(1 + \frac{4}{\pi} \OmegaE^{-3}\right) \right]\,.
    \label{eq:gap3S1++}
\end{align}

\section{Comparison between solutions of the RG equation and the gap equation}
\label{sec:gapeq}

Let us now focus on the leading coefficient in the perturbative series of $\ln(\Delta/\mu)$, and compare these values in the gap equation and the preceding RG equation.
We also discuss the overall coefficient of $\Delta$.

\subsection{Leading coefficient of $\ln(\Delta/\mu)$}

The gap equation in $J=0$ channel is
\begin{align}
    \Delta (p_0)
    &= \frac{g^2}{12\pi^2} \int dq_0 \int d \theta\, \sin\theta \bigg(\frac{\frac12 + \frac12 \cos\theta}{1 - \cos\theta + \left[\Pi^{\mathrm{E}} + (p_0 - q_0)^2\right]/(2\mu^2)} \label{eq:gapeqJ0}\\
    &\qquad \qquad \qquad \qquad \qquad \quad + \frac{\frac32 - \frac12 \cos\theta}{1 - \cos\theta + \left[\Pi^{\mathrm{M}} + (p_0 - q_0)^2\right]/(2\mu^2)}
    \bigg)
    \frac{\Delta(q_0)}{\sqrt{q_0^2 + \Delta(q_0)^2}}\,.\notag
\end{align}
The $J=0$ channel contains ${}^1S_0$ and ${}^3P_0$ channels, and we will exclusively extract the ${}^1S_0$ contribution below.
One-gluon exchange amplitude in Eq.~\eqref{eq:M++} can be rewritten as
\begin{align}
    \calM_{++} (\theta)
    = G^0_{++} \frac{1 + \cos\theta}{2} + G^i_{++} \frac{3 - \cos\theta}{2}\,.
    \label{eq:M++1}
\end{align}
By contrasting this expression with the gap equation, one can clearly see that the chromoelectric and magnetic gluon propagators in the bracket correspond to $G^0_{++}$ and $G^i_{++}$, respectively.
One can further rewrite the one-gluon exchange amplitude as
\begin{align}
    \calM_{++} (\theta)
    &= \frac12 \left(G^0_{++} + 3 G^i_{++}\right) + \frac12 \left(G^0_{++} - G^i_{++}\right)\cos\theta\,.
    \label{eq:M++2}
\end{align}
From Eq.~\eqref{eq:Hsinglet}, the helicity amplitude in the ${}^1S_0$ channel is
\begin{align}
    \mathcal{H}_{++}^{{}^1S_0} &= G^0_{0++} + 3G^i_{0++}\,,
\end{align}
and from Eq.~\eqref{eq:Htriplet}, the helicity amplitude in the ${}^3P_0$ channel is
\begin{align}
    \mathcal{H}_{++}^{{}^3P_0} &= G^0_{1++} - G^i_{1++}\,.
\end{align}
As is evident from the comparison between these helicity amplitudes and the one-gluon exchange amplitude~\eqref{eq:M++2}, one can clearly see that the ${}^1S_0$ amplitude corresponds to the $\cos\theta$-independent term in Eq.~\eqref{eq:M++2}.
Therefore, to extract the $^1S_0$ component in the gap equation, we drop the $\cos\theta$-dependent term in the gluon propagators.
The resulting gap equation is
\begin{align}
    \Delta (p_0)
    &= \frac{g^2}{12\pi^2} \int dq_0 \int d \theta\, \sin\theta \bigg(\frac{1/2}{1 - \cos\theta + \left[\Pi^{\mathrm{E}} + (p_0 - q_0)^2\right]/(2\mu^2)} \label{eq:gapeq1S0}\\
    &\qquad \qquad \qquad \qquad \qquad \quad + \frac{3/2}{1 - \cos\theta + \left[\Pi^{\mathrm{M}} + (p_0 - q_0)^2\right]/(2\mu^2)}
    \bigg)
    \frac{\Delta(q_0)}{\sqrt{q_0^2 + \Delta(q_0)^2}}\,.\notag
\end{align}
By substituting the expression for $\Pi^{\mathrm{M}}$ and carrying out the $\theta$-integral, we find to the leading logarithm
\begin{align}
    \Delta (p_0)
    &= \frac{g^2}{12\pi^2} \int dq_0 \ln \left(\frac{\mu}{|p_0 - q_0|}\right)
    \frac{\Delta(q_0)}{\sqrt{q_0^2 + \Delta(q_0)^2}}\,.
    \label{eq:gapeq1S0LL}
\end{align}
One can solve this by using the method in Refs.~\cite{Son:1998uk,Schafer:1999jg}.
The solution of this gives the leading coefficient
\begin{align}
    \ln \left(\frac{\Delta_0}{\mu}\right) \simeq - \frac{\sqrt{3}\pi^2}{g}\,.
\end{align}
This agrees with the solution of our RG equation.
This coefficient is $\sqrt{2/3}$ times that of the solution of the gap equation in $J=0$ channel.
This is owing to the difference in the coefficient of the magnetic gluon propagator.
The $\theta$-integral in the gap equation is dominant in the forward region $\theta \simeq 0$, so the coefficient of the magnetic gluon propagator in Eq.~\eqref{eq:gapeqJ0} is $3/2 - \cos\theta /2 \simeq 1$ while the coefficient of that in Eq.~\eqref{eq:gapeq1S0} is $3/2$.
So, in $J=0$ case, the prefactor in Eq.~\eqref{eq:gapeq1S0LL} is $g^2/(18\pi^2)$, instead of $g^2 /(12\pi^2)$ in ${}^1S_0$ case.

Let us also compare our result with Son's RG equation~\cite{Son:1998uk}.
Son's RG equation only takes into account the magnetic sector.
The electric sector also has an attraction; they are screened at a small angle, but they also contribute to the pairing through a large angle scattering.
Neglecting the self-energy correction, as it was not taken into account in Ref.~\cite{Son:1998uk}, the RG equation with only the magnetic sector can be found by setting $G^0_{L++}=0$ in Eq.~\eqref{eq:RGVs++}:
\begin{align}
    \frac{d G^i_{L++}}{dt} = - \frac{3}{2} N \left(G^i_{L++}\right)^2 + \frac{c_R g^2}{6\mu^2}\,.
\end{align}
On the other hand, Son's RG equation reads
\begin{align}
    \frac{d G^i_{L++}}{dt} = -N \left(G^i_{L++}\right)^2 + \frac{c_R g^2}{6\mu^2}\,.
\end{align}
Again, the factor $3/2$ difference in the coefficient of the one-loop beta function resulted in the factor $\sqrt{2/3}$ difference in the leading expansion term of $\ln(\Delta/\mu)$.
This additional factor $3/2$ arises from the combination of the electric and magnetic coupling functions.
One cannot neglect the electric sector even though the magnetic sector is dominant.

\subsection{Full perturbative series up to $O(g^0)$}
\begin{figure}
    \centering
    \includegraphics[width=0.5\linewidth]{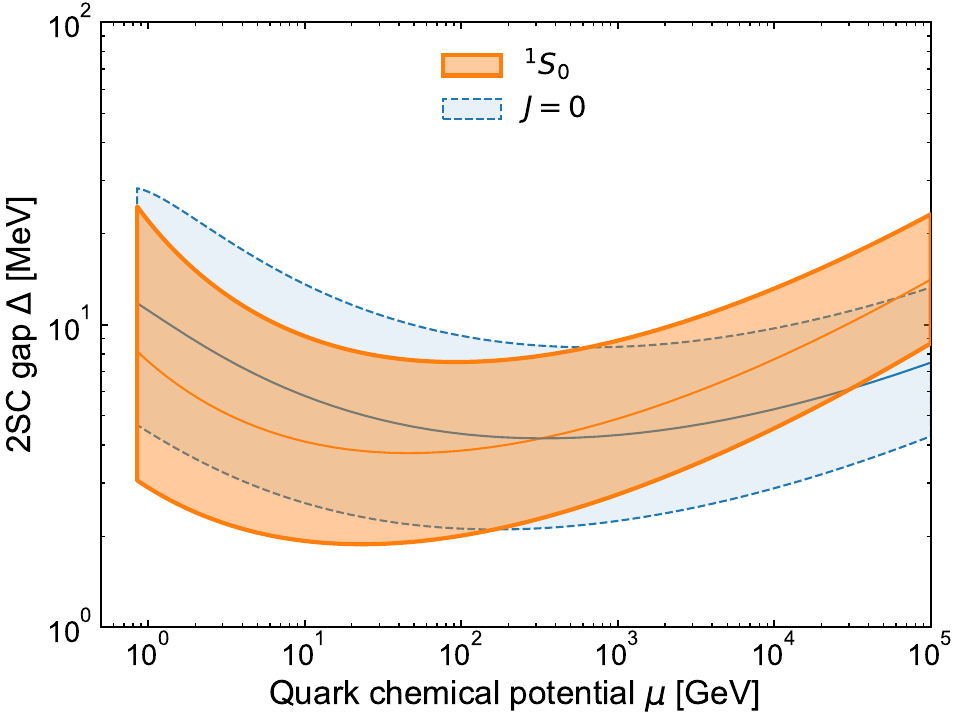}
    \caption{The 2SC gap in the $^1S_0$ channel. The gap in the $J=0$ channel is overlaid for comparison. Overall, the $^1 S_0$ channel has larger gap because it has stronger attraction compared to the $J=0$ channel in which the repulsive $^3P_0$ channel is coupled.}
    \label{fig:J0_1S0}
\end{figure}

Let us now reproduce the solution of the gap equation in the $J=0$ channel, which we denote it as $J=0$ for brevity, from the solution of the RG equation in the ${}^1S_0$ channel we discussed above~\eqref{eq:gap1S0}.
The $J=0$ gap reads
\begin{align}
    \ln\left(\frac{\Delta_{J=0}}{\mu}\right) = - \frac{3\pi^2}{\sqrt{2}g} - \frac{\pi^2 + 4}{8} + \ln\left(\frac{16}{\pi}\OmegaE^{-5}\right)\,.
    \label{eq:gapJ0}
\end{align}
Instead of Eq.~\eqref{eq:gluonprop}, if we use the following gluon propagator in the matching condition, we can reproduce the $J=0$ result:
\begin{align}
    \DE &= \frac{1 + \cos\theta}{2\mu^2(1-\cos\theta) + \Pi^{\mathrm{E}}}\,,\\
    \DM &= \frac{1 - \frac13 \cos\theta}{2\mu^2(1-\cos\theta) + \Pi^{\mathrm{M}}}\,.
\end{align}
Note that the numerator of the propagator is replaced to match with gluon propagators in Eq.~\eqref{eq:gapeqJ0}.
The gap from this matching condition is
\begin{align}
    \ln\left(\frac{\Delta_{J=0,\mathrm{RG}}}{\mu}\right) = - \frac{3\pi^2}{\sqrt{2}g} - \frac{\pi^2 + 4}{8} + \ln\left[2\OmegaE \left(1 + \OmegaE^{-2}\right)^{\frac32} \left(1 + \frac{4}{\pi} \OmegaE^{-3}\right) \right]\,.
    \label{eq:gapJ0RG}
\end{align}
Precisely speaking, such a derivation is inconsistent in terms of the angular momentum structure because one has to consider the coupled-RG equation between $^1S_0$ and $^3P_0$ channels as they are mixed in the $J=0$ channel.
Nevertheless, this comparison provides the cross check with the existing calculation.

In Fig.~\ref{fig:J0_1S0}, we compare the values of the $J=0$ gap and the $^1S_0$ gap in the 2SC phase.
As in Fig.~\ref{fig:gap_compare}, the band corresponds to the scale variation in the running coupling.
In $J=0$, there is also repulsive $^3P_0$ channel is contained, and the leading coefficient is smaller compared to the pure $^1S_0$ channel, so it is natural to expect that the gap is smaller compared to the pure $^1S_0$, but this is not the case as one can see in Fig.~\ref{fig:J0_1S0}.

Such difference arises from the difference in the electric and magnetic propagator integrated from small to large angles.
The ratio of the coefficients of the electric and magnetic sector is $1:3$ in the $^1S_0$, while it is $1:1$ in the $J=0$ channel, so the electric contribution is three times larger in the $J=0$ case compared to $^1 S_0$ case.
In the $^1S_0$ gap formula \eqref{eq:gap1S0}, the term arising from the angular integral of the electric and magnetic propagator is 
\begin{align}
    \ln\left[2\OmegaE\left(1 + \OmegaE^{-2}\right)^{\frac12} \left(1 + \frac{4}{\pi} \OmegaE^{-3}\right) \right]\,.
\end{align}
The term $(1 + \OmegaE^{-2})^{\frac12}$ is from the electric and $(1 + \frac{4}{\pi} \OmegaE^{-3})$ is from the magnetic sector.
To compare with the $J=0$ result, we make the power of the electric factor three times larger, namely,
\begin{align}
    \ln\left[2\OmegaE\left(1 + \OmegaE^{-2}\right)^{\frac32} \left(1 + \frac{4}{\pi} \OmegaE^{-3}\right) \right] \simeq \ln\left(\frac{8}{\pi}\OmegaE^{-5} \right)\,.
    \label{eq:logJ0}
\end{align}
On the right-hand side, we neglected the factor one in the brackets.
The right-hand side of this equation corresponds to the last term in the $J=0$ gap \eqref{eq:gapJ0}.
This deficit in the electric sector by a factor $\OmegaE^{-2}$ in the $^1S_0$ channel makes the gap smaller compared to the $J=0$ case.

\subsection{Overall coefficient in $\Delta$}

Now, by comparing the third term in Eq.~\eqref{eq:gapJ0} and Eq.~\eqref{eq:logJ0}, we observe the factor two difference in the argument of the logarithm.
This is related to ambiguity in the overall coefficient of the gap;
this overall coefficient is generally known to have ambiguity up to a factor of order one.
Indeed, in the earliest literature, the overall coefficient was $256 \pi^4 g^{-5}$ ($\Nf=2$)~\cite{Schafer:1999jg}, but the oft-quoted value now is twice as large $512 \pi^4 g^{-5}$, which is also the value adopted in Eq.~\eqref{eq:gapeqJ0}.

We claim that such ambiguity of the overall coefficient in the gap is absent in the RG evaluation.
Let us illustrate this.
This ambiguity in the RG equation corresponds to the scale at which we impose the matching condition with perturbation theory.
The matching should be imposed at the soft scale $\Lambda \sim \mD$, but its specific value can be $\Lambda = X \mD$, where $X$ is a factor of order one.
Then, by choosing $\Lambda = X\mD$, the gap \eqref{eq:gap1S0} becomes
\begin{align}
    \ln\left(\frac{\Delta}{\mu}\right)
    &= - \frac{\sqrt{3}\pi^2}{g} - \frac{\pi^2 + 4}{12} + \ln\left[2X\OmegaE\left(1 + \OmegaE^{-2}\right)^{\frac12} \left(1 + \frac{4}{X \pi} \OmegaE^{-3}\right) \right]\,,\notag\\
    & \simeq - \frac{\sqrt{3}\pi^2}{g} - \frac{\pi^2 + 4}{12} + \ln\left(\frac{8}{\pi}\OmegaE^{-3}\right)\,. 
\end{align}
As one can see, the $X$ dependence cancels in the last expression.
Therefore, the RG equation can set the coefficient of the gap unambiguously.

Given the absence of the ambiguity of the overall coefficient, one can compare the $J=0$ coefficient, which is the last term in \eqref{eq:gapJ0}, with Eq.~\eqref{eq:logJ0}.
From the comparison, we conclude that the oft-quoted overall coefficient is overestimated by a factor two.
That is, the last term of Eq.~\eqref{eq:gapJ0} should be $\ln[(8/\pi)\OmegaE^{-5}]$ instead of $\ln[(16/\pi)\OmegaE^{-5}]$, so the overall coefficient of the $J=0$ gap may be small by a factor two.

\section{Superfluid gap in QCD at finite isospin density}
\label{sec:isospin}

In this section, we evaluate the superfluid gap in QCD at finite \emph{isospin} density, in which one assigns the isospin chemical potential $\mu_I/2$ to $u$ quark, and $-\mu_I/2$ to $d$ quark.
So far, this is the only theory in which the lattice-QCD calculation up to large chemical potential is available~\cite{Abbott:2023coj,Abbott:2024vhj} so that the direct comparison with the weak-coupling calculation is feasible~\cite{Fujimoto:2023mvc,Fujimoto:2024pcd}.
In this theory, apart from few minor modifications in the derivation, the same RG equation holds as in the case of QCD at finite baryon density.
The largest modification is that the pairing in this theory is not in a diquark $qq$ channel but in a $\bar{q}q$ channel, so the color coefficient is $c_R = c_F = - (\Nc^2 - 1) / 2\Nc$.
Because the one-gluon attraction is stronger in $\bar{q}q$ channel compared to the diquark channel, the magnitude of the gap is enhanced at finite isospin density.
The pairing in this theory is among different flavors, so only the $t$-channel gluon exchange contributes, while the $s$-channel diagram does not.

This large gap will affect thermodynamic quantities through the Cooper pair condensation energy $\propto \mu^2 \Delta^2$.
Initially, the gap seemed to be large in the lattice calculation~\cite{Abbott:2023coj}, but the latest calculation in the continuum limit suggests that the gap is overestimated in weak-coupling calculation~\cite{Abbott:2024vhj}.

\begin{figure}
    \centering
    \includegraphics[width=0.5\linewidth]{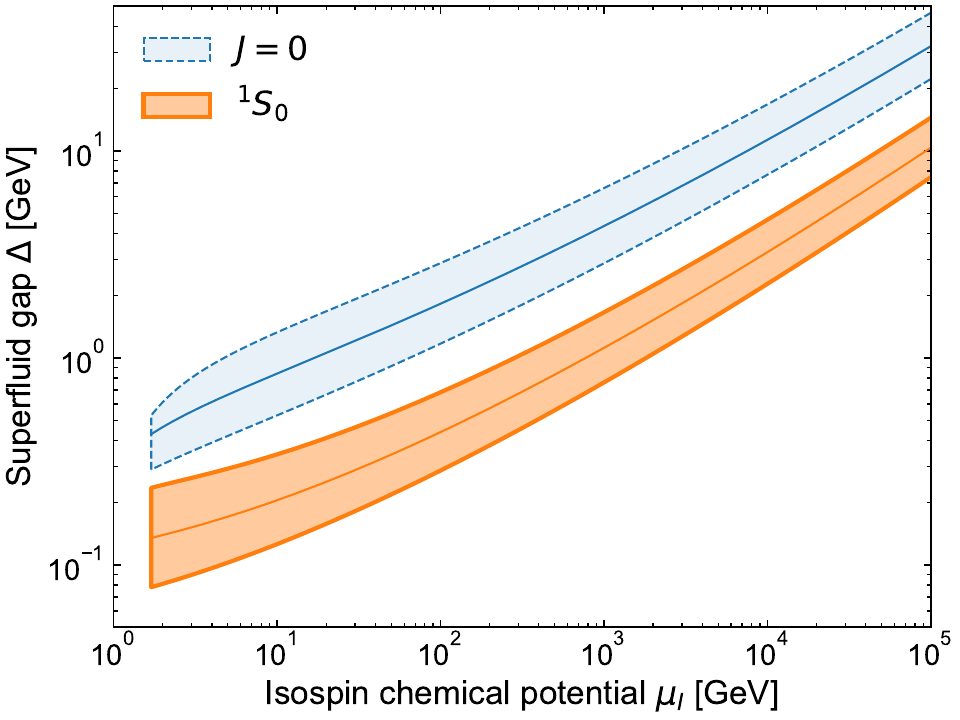}
    \caption{Comparison of the superfluid gap in the weak-coupling regime of QCD at finite isospin density evaluated from the RG equation (${}^1S_0$) and from the gap equation ($J=0$). See the text for detailed explanation.}
    \label{fig:piongap}
\end{figure}
By substituting the following parameters in Eq.~\eqref{eq:lngap}
\begin{align*}
    a = \frac12\,,\quad
    b = \frac13\,,\quad
    l = \ln\left[\left(1 + \OmegaE^{-2}\right)^{\frac12} \left(1 + \frac{4}{\pi} \OmegaE^{-3}\right) \right]\,,
\end{align*}
we can evaluate the superfluid gap as
\begin{align}
    \ln\left(\frac{\Delta_{{}^1S_0}}{\mu}\right) = - \frac{\sqrt{3}\pi^2}{\sqrt{2}g} - \frac{\pi^2 + 4}{24} + \ln\left[2\OmegaE \left(1 + \OmegaE^{-2}\right)^{\frac12} \left(1 + \frac{4}{\pi} \OmegaE^{-3}\right) \right]\,.
\end{align}
On the other hand, one can also evaluate the gap in the $J=0$ channel as a solution of the gap equation:
\begin{align}
    \ln\left(\frac{\Delta_{J=0}}{\mu}\right) = - \frac{3\pi^2}{2g} - \frac{\pi^2 + 4}{16} + \ln\left(\frac{16}{\pi}\OmegaE^{-5}\right)\,.
\end{align}
We show these values of superfluid gap as a function of isospin chemical potential $|\mu_I| \equiv 2\mu$ in Fig.~\ref{fig:piongap}.
Again, we use the running coupling constant from the two-loop QCD beta function but with $\Nf = 3$ in the vacuum loops because these lattice data are evaluated on the $\Nf = 3$ ensemble.
Note that $\Nf = 2$ is used in the other part of the calculation, e.g., in the Debye mass.
The band corresponds to the scale variation.
The gap is larger in the $J=0$ channel compared to the $^1S_0$ channel for the same reason mentioned in the previous section;
this is owing to the difference in the electric sector of the gluon propagator.
The ${}^1S_0$ gap becomes larger only at $\mu_I \gtrsim 10^{14}\,\text{GeV}$.

\begin{figure}
    \centering
    \includegraphics[width=0.48\linewidth]{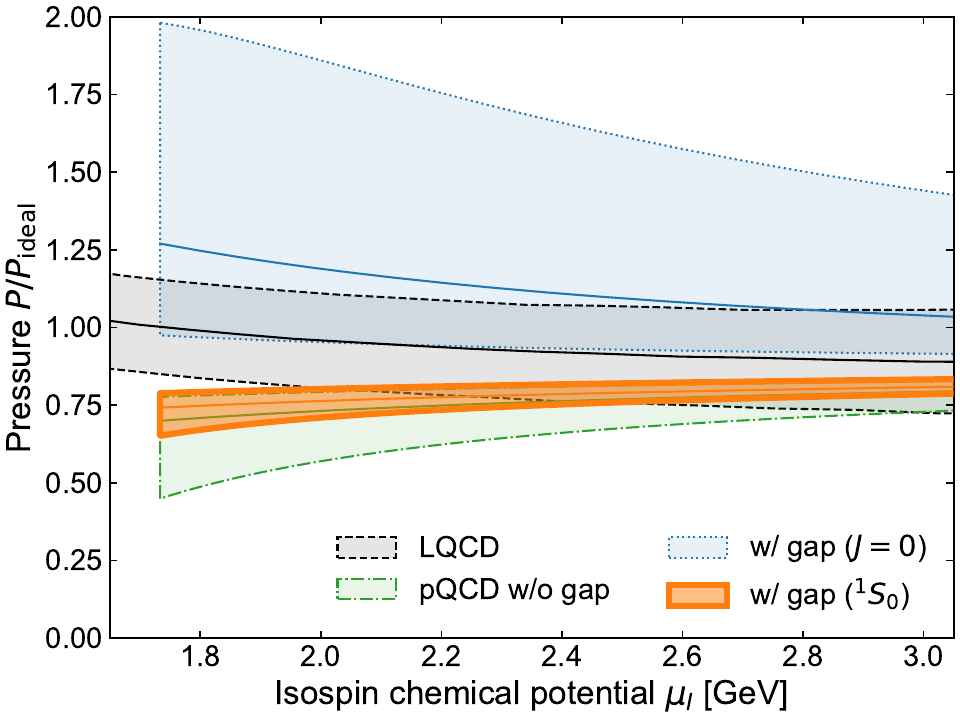}\hfill
    \includegraphics[width=0.48\linewidth]{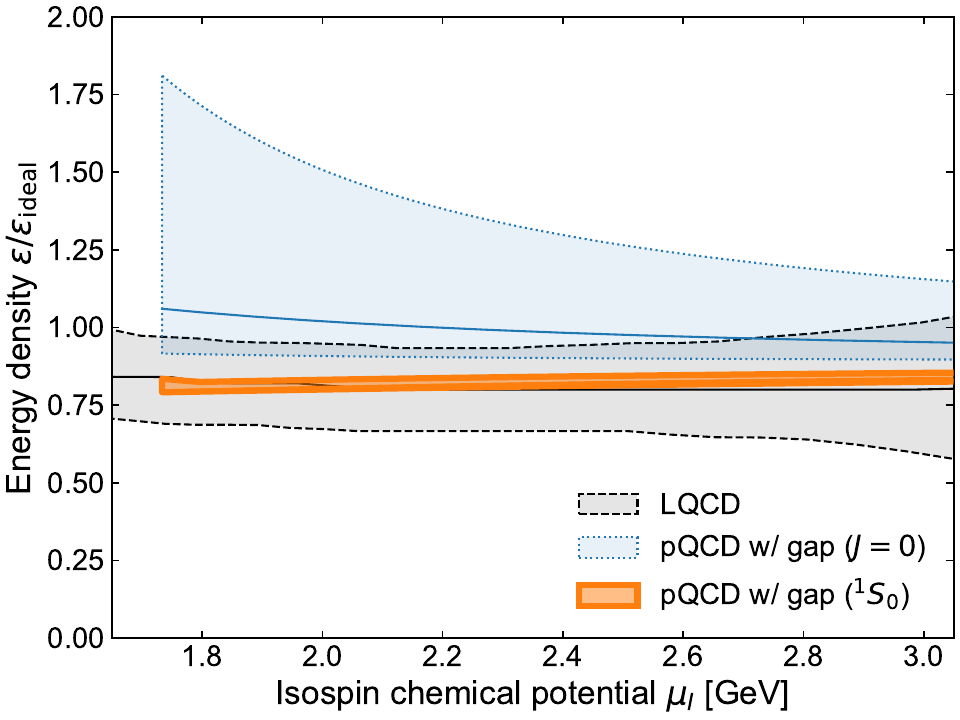}\hfill
    \includegraphics[width=0.48\linewidth]{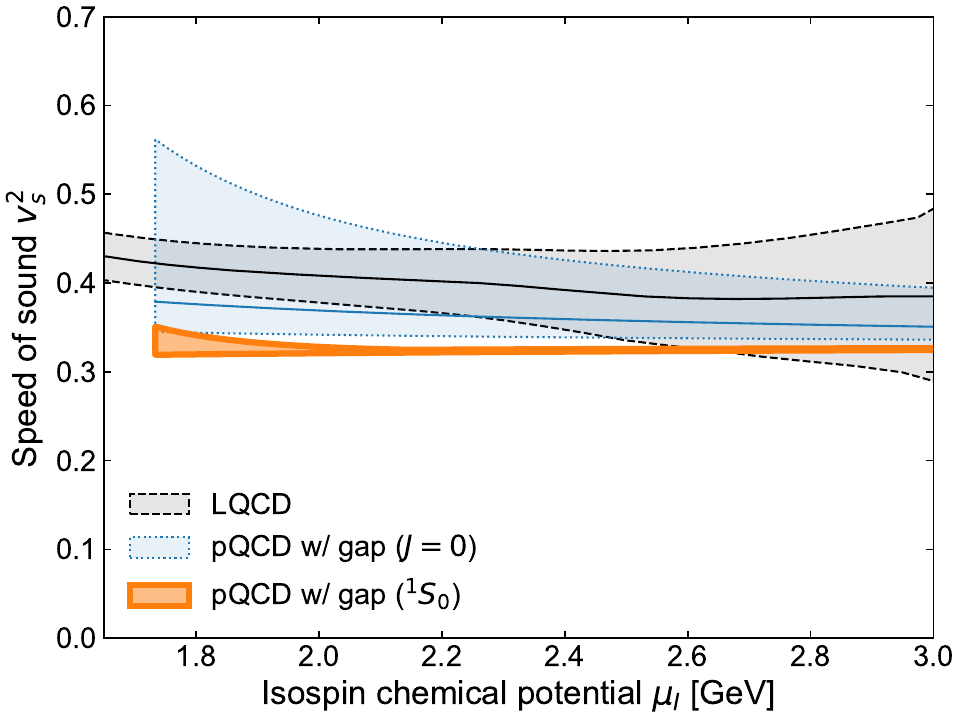}
    \caption{The weak-coupling calculation of thermodynamic quantities with the condensation energy contribution compared with the lattice-QCD thermodynamics. Center line in the shaded region corresponds to the scale choice $\bar{\Lambda} = \mu_I$, and the band corresponds to the scale variation between $\mu_I/2 < \bar{\Lambda} < 2\mu_I$. \textit{Top left:} Pressure normalized by the ideal quark gas pressure $P_{\mathrm{ideal}} = \Nc \Nf \mu^4 / (12\pi^2)$. \textit{Top right:} Energy density normalized by the ideal quark gas value $\varepsilon_{\mathrm{ideal}} = \Nc \Nf \mu^4 / (12\pi^2)$. \textit{Bottom:} Speed of sound.}
    \label{fig:QCDI}
\end{figure}

In Fig.~\ref{fig:QCDI}, we compare the weak-coupling calculation of the thermodynamics and the lattice-QCD data.
We contrast the previous estimate with the $J=0$ gap and the current estimate with the $^1S_0$ gap.
The pressure in perturbation theory is
\begin{align}
    \frac{P}{P_{\mathrm{ideal}}} = 1 - 2\frac{\alpha_s}{\pi} - \left[\Nf \ln \left(\Nf \frac{\alpha_s}{\pi}\right) + \beta_0\ln \left(\frac{\bar{\Lambda}}{2\mu}\right) + \frac{67}3 + \frac{11}3 \beta_0 + 0.224 \Nf \right]
    \bigg(\frac{\alpha_s}{\pi}\bigg)^2\,,
    \label{eq:PpQCD}
\end{align}
where $\beta_0$ is the one-loop coefficient of the QCD beta function, and the pressure of the ideal quark gas $P_{\mathrm{ideal}}$ is defined as $P_{\mathrm{ideal}} = \Nc \Nf \mu^4 / (12\pi^2)$.
The condensation energy is
\begin{align}
    \delta P = \frac{3\mu^2 \Delta^2}{2\pi^2} \left(1 - \frac{b_{-1}}{9\pi^2} g + \cdots \right) \,,
\end{align}
where $b_{-1}$ is the coefficient of the $O(1/g)$ term in $\ln(\Delta / \mu)$.
For the details of the $\delta P$ contribution to thermodynamics, readers can refer to Ref.~\cite{Fujimoto:2023mvc}.

The effect of the gap seems to be overestimated in the $J=0$ calculation in the pressure and energy density, while it is smaller in the $^1S_0$ calculation, and is slightly better explained.
The behavior of the speed of sound favors the $J=0$ gap over the $^1S_0$ gap.
Given these observations, at the moment, we cannot draw the conclusive statement which gap function is favored from these comparisons as neither of them aligns perfectly with all the lattice-QCD data.
Nevertheless, the thermodynamics should lie somewhere between these two evaluations, which seems to be the case from these figures.

It may also be possible that the weak-coupling calculation is invalid around $\mu_I\simeq 2\,\text{GeV}$, and is only valid at even larger $\mu_I$, or there are some other hidden sources of discrepancy than the gap contribution.
In particular, the $\mu_I \simeq 2\,\text{GeV}$ corresponds to the baryon chemical potential of $\mu_B \simeq 3\,\text{GeV}$, and this is the value down to which the weak-coupling calculation is used in neutron star phenomenology (see Ref.~\cite{Kurkela:2014vha} for example), so testing whether the weak coupling calculation is important.
To this end, the precise determination of the gap parameter is necessary.

\section{Discussions}
\label{sec:discussions}

In this section, we mention two origins that can either suppress or enhance the magnitude of the gap.
Finally, we close this section by mentioning the convergence and the possible extension to higher order.

\subsection{Complication associated with the Meissner screening}

We turn to the complication in the secondary gap. 
The secondary pairing gap is among quarks that is left ungapped in the primary diquark pairing.
Let us consider an example of the 2SC phase in $\Nf =2$ case.
In the 2SC phase, only quarks with two out of three colors (e.g. red and green) undergo pairing in the gauge-fixed description, and quarks with the remaining color (e.g. blue) do not.
The remaining blue diquarks are in the color $\boldsymbol{6}$ representation, which has been thought to have repulsive one-gluon exchange interaction.
However, as we have discussed above, this channel has an attraction and can form the Cooper pair (see cases IV and V in Sec.~\ref{sec:classify}).

In the 2SC phase, the primary diquark condensate, which can also be regarded as a Higgs condensate, can ``break'' the gauge redundancy as $SU(3)_c \to SU(2)_c$.
The unbroken generator of $SU(2)_c$ mediates the interaction between the red and green quarks.
The secondary pairing among blue quarks are mediated by the gluons corresponding to the broken generators, so the interaction is screened as it is subject to the color Meissner effect.
This color Meissner effect is the source of the complication as we will explain below.

Let the gap of the primary pairing be $\Delta_1$.
Then, the secondary pairing takes place at the energy scale $q_0$ below the primary pairing gap $q_0 \ll \Delta_1$.
At this energy scale, the magnetic interaction between the blue quarks are described by the gluon propagator~\eqref{eq:gluonprop} with the self-energy~\cite{Rischke:2000qz,Rischke:2002rz}
\begin{equation}
    \Pi^{\mathrm{M}}(q_0, q) = \frac{\pi^2}{8} \mD^2 \frac{\Delta_1}{q}\,, \quad (q_0 \ll \Delta_1 \ll q)\,.
\end{equation}
This indicates that the screening becomes static, i.e., it does not depend on the energy $q_0$ of the ultra-soft scale, so the tree-level beta function does not appear in the RG equation.
The tree-level beta function \eqref{eq:betatree} in the RG equation \eqref{eq:RGschematic} should be modified as
\begin{align}
    \tilde{\beta}^i_{\mathrm{tree}}(t)
    =
    \begin{cases}
        \dfrac{c_R g^2}{6\mu^2} & (0 < t < t_1)\,,\\
        0 & (t> t_1)\,,
    \end{cases}
\end{align}
where $t_1 = - \ln (\Delta_1 / \mD)$ is the scale parameter associated with the energy scale at which the primary pairing occurs.

This Meissner screening may lead to the following.
The $\beta$ function associated with the renormalization of the tree-level amplitude in the RG equation now gets modified as:
\begin{equation}
    \tilde{\beta}_{\mathrm{tree}}(t) =
    \begin{cases}
        \beta_{\mathrm{tree}} & 0 < t < t_1\,, \\
        0 & t > t_1\,.
    \end{cases}
\end{equation}
where we defined as $t_1 \equiv -\ln(\Delta_1 / \mD)$.
This means that we solve this RG equation from $t=0$ to $t_1$:
\begin{align}
    \frac{d \Vtrip_{L++}}{dt} &= -\frac{\left(\Vtrip_{L++}\right)^2}{6\left(1 + \frac{g^2}{9 \pi^2}t \right)}  - \frac{g^2}{36\pi^2}\,,
    \label{eq:RGMeissner1}
\end{align}
and then from $t=t_1$ to $t_{\mathrm{BCS}}$, namely, up to the scale at which the solution of the following RG equation hits the singularity:
\begin{align}
    \frac{d \Vtrip_{L++}}{dt} &= -\frac{\left(\Vtrip_{L++}\right)^2}{6\left(1 + \frac{g^2}{9 \pi^2}t \right)}\,.
\end{align}
From this solution, the secondary gap $\Delta_2$ can be read out as
\begin{align}
    \ln \left(\frac{\Delta_2}{\Delta_1}\right) \simeq - \frac{6}{|\Vtrip_{L++}(t=t_1)|} - \frac{2g^2}{\pi^2|\Vtrip_{L++}(t=t_1)|^2}\,,
\end{align}
where $\Vtrip_{L++}(t=t_1)$ is obtained by integrating Eq.~\eqref{eq:RGMeissner1} from $t=0$ to $t_1$.
The concrete expression can be found in Appendix~\ref{sec:solRGeq}.

As we have seen in Fig.~\ref{fig:gap_compare}, the secondary pairing in the color $\boldsymbol{6}_c$ is already too small to be phenomenologically relevant.
The Meissner effect may further suppress the magnitude of the gap particularly at high density.

\subsection{Instanton effect}

Let us turn to another effect that can possibly amplify the magnitude of gap particularly at low density.
This is the instanton effect.
Below we sketch how this might lead to an enhancement.
The interaction among massless quarks mediated by instanton is
\begin{align}
    \mathcal{L}_{\mathrm{instanton}} = K^{\mathrm{S}} \det_f (\bar{\psi}_+ \psi_-) - K^{\mathrm{T}} \det_f (\bar{\psi}_+ \sigma^{\mu\nu}\psi_-)\,,
    \label{eq:Linst}
\end{align}
where $+$ and $-$ refer to the chirality of quarks, which matches with helicity in the massless limit.
This interaction is flavor-singlet.
Roughly speaking, the RG equation becomes
\begin{align}
    \frac{d \left(\tilde{G} \pm \tilde{K}\right)}{dt} & = - \gamma \frac{N\left(\tilde{G} \pm \tilde{K}\right)^2}{\left(1 + \frac{g^2}{9\pi^2}t\right)} + \alpha \beta^i_{\mathrm{tree}}\,,
\end{align}
where $\tilde{G} = G^0 + \alpha G^i$ collectively denotes the singlet or triplet combination of the electric and magnetic coupling functions, and $\tilde{K} = K^{\mathrm{S}} - K^{\mathrm{T}}$.
The coefficient $\gamma$ a factor that depends on the helicity and spin.
One of the equation gives the enhanced gap, depending on the sign of $\tilde{K}$, because the net attraction is increased by the factor $\pm \tilde{K}$.
We also note that the quark operator in the instanton interaction~\eqref{eq:Linst} is flavor singlet, so this can only contribute to the case I and V in the classification in Sec.~\ref{sec:classify}.

\subsection{Convergence and road map to the higher-order corrections}

Finally, let us discuss the details related to the perturbative series of $\ln(\Delta/\mu)$, namely, its convergence and the possible road map to the higher-order corrections.

By substituting the values in Eq.~\eqref{eq:gap1S0} for the $\Nf=2$ case, one can evaluate the actual numbers as
\begin{align}
    \ln\left(\frac{\Delta}{\mu}\right) \simeq -\frac{\sqrt{3}\pi}{2} \left(\frac{\alpha_s}{\pi}\right)^{-\frac12} \left[1 + 0.866 \left(\frac{\alpha_s}{\pi}\right)^{\frac12} + 0.511 \left(\frac{\alpha_s}{\pi}\right)^{\frac12} \ln \left(\frac{\alpha_s}{\pi}\right) + \cdots\right]\,.
\end{align}
The coefficient of each term is smaller than unity.
This is the lowest order correction, so one cannot determine conclusively whether this series is converging well or not, nevertheless, a radius of convergence seems to be reasonable even compared to other perturbative series in finite-density QCD, see Eq.~\eqref{eq:PpQCD} for example.
That said, the leading correction starts at $O(\alpha_s^{1/2})$, and the higher-order corrections are still at $O(\alpha_s)$, so they are likely to be important.

In the calculation of the gap, there are two hierarchy of organizing the series expansion.
The one is in terms of the ordinary expansion in terms of the coupling constant $g$.
The other is whether the beta function is relevant in the sense of the RG transformation.
In principle, the irrelevant beta function can also modify the gap (see, e.g., Ref.~\cite{Fujimoto:2025zfa}).
However, if the leading-order coefficient dominates the gap, corrections from the irrelevant beta functions in the RG equation turns out to be negligible.
This is confirmed by numerically solving the RG equation contrasting two cases with and without the irrelevant beta functions.
Particularly, when there is a constant term in the RG equation arising from the tree-level renormalization, the exponentially decaying irrelevant beta function are totally negligible.

The $t$-dependent term of the RG equations (\ref{eq:RGVs++}, \ref{eq:RGVt++}, \ref{eq:RGVt+-}) should be unchanged with the higher order corrections because there are no new beta function coming in at higher order as we already explained above.
So, the form of the solution \eqref{eq:lngap} remains the same when going higher order in $g$, and the only there are higher-order corrections to the parameters $a$, $b$, and $L$.
So, the only higher order correction comes in through the matching condition between the effective theory and QCD.
All one has to consider is just the one-loop correction to the tree-level diagram in HDL effective theory.
The technologies has already been developed in Refs.~\cite{Braaten:1989mz,Bellac:2011kqa,Gorda:2021kme}.
Such computations of the higher-order correction are under way.

\section{Conclusions and outlook}
\label{sec:summary}

In this work, we have systematically revisited the calculation of the color-superconducting gap using the renormalization group (RG) approach.
Within this framework, Son's beta function~\cite{Son:1998uk} emerges naturally, and the quark self-energy corrections can be consistently understood in the context of the hard dense loop (HDL) effective theory.
We have shown that the RG method reproduces, at the same perturbative order, the gap previously obtained by other approaches, and---crucially, as discussed at the end of Sec.~\ref{sec:gapeq}---allows the coefficient of the gap to be fixed unambiguously.

Traditionally, one-loop gap calculations have relied on solving the gap equation~\cite{Son:1998uk, Pisarski:1998nh, Pisarski:1999av, Pisarski:1999bf, Pisarski:1999tv, Schafer:1999jg, Hong:1998tn, Hong:1999ru,Hong:1999fh, Wang:2001aq} or summing ladder diagrams in the two-particle correlation function to extract the singularity~\cite{Brown:1999aq, Brown:1999yd, Brown:2000eh}.
In our view, the RG equation offers the simplest route to this computation.
Whereas perturbation theory requires explicit summation of ladder diagrams as higher-order corrections, the RG framework performs this summation automatically via the solution of a differential equation.
Perturbative corrections enter only through the matching conditions between the effective theory and QCD, which reduce to tree-level quark-quark scattering diagrams.
This separation between perturbative corrections and ladder resummation significantly streamlines the analysis, and in the present case, the RG equation admits an analytic solution [Eq.~\eqref{eq:generalsol}].
While higher-order corrections are challenging in other approaches, here they reduce to evaluating corrections to tree-level diagrams within the HDL effective theory under restricted kinematics.

Higher-order perturbative corrections are not merely of academic interest:
they are essential for improving the convergence of the expansion and reducing the renormalization-scale dependence.
Since the leading perturbative correction is of order $O(g)$ rather than $O(g^2)$, its quantitative impact is substantial.
From the perspective of applications to the QCD phase diagram and the equation of state, this improvement is highly relevant.
In the phase diagram, the competition between the stress induced by finite quark masses (which tend to suppress the gap) and the gap's compensating effect is delicate.
As discussed in the accompanying paper~\cite{Fujimoto:2025liq}, reducing the uncertainty in the gap allows for a more precise determination of the weak-coupling phase boundaries, which in turn constrains the nonperturbative, small-$\mu$ region relevant for neutron-star physics.
From the equation-of-state perspective, it also facilitates more meaningful cross-checks between lattice QCD at finite isospin density and weak-coupling QCD results.
Although current comparisons rely on accidental cancellations of scale dependence between the gap and bulk thermodynamic quantities, as seen in Fig.~\ref{fig:QCDI}, the present approach may enable more unambiguous matching in the future.

The RG equations naturally decompose into independent channels labeled by color, flavor, helicity, and angular-momentum term symbols.
While our analysis has been restricted to the massless limit, finite quark masses induce mixing between different helicity channels.
In the $u$-$d$ sector, such effects are expected to be small;
however, for strange quarks with masses of order $100 \,\text{MeV}$, the mixing could be significant.
The computation of these mass effects, along with related higher-order corrections, is currently in progress and will be reported elsewhere.

\begin{acknowledgments}
This research was supported in part by the Japan Science and Technology Agency (JST) as part of Adopting Sustainable Partnerships for Innovative Research Ecosystem (ASPIRE) Grant No.\ JPMJAP2318, the National Science Foundation Grant No.\ PHY-2020275, and the Heising-Simons Foundation Grant 2017-228.
\end{acknowledgments}

\appendix

\section{Details of the calculation}

\subsection{One-loop beta function}
\label{sec:IT}

In this subsection, we show detailed calculation for Eqs.~(\ref{eq:calI}, \ref{eq:calT}) that leads to the one-loop RG equations (\ref{eq:RG0++}, \ref{eq:RGi++}, \ref{eq:RG0+-}, \ref{eq:RGi+-}):
\begin{align}
    \tag{\ref{eq:calI}}
    \mathcal{I} &\equiv -\frac14 \int_{-\infty}^{\infty} \frac{dk_0}{2\pi} \int_{d\Lambda} \frac{k^2 \, d k}{2\pi^2} \,
    \frac{1}{k_0^2 - \epsilon_k^2}\,, \\
    \mathcal{T}_{\lambda_1 \lambda_2}
    &\equiv
    \int \frac{d\hat{k}}{4\pi} \Big\{
    G_{\mu\nu;\lambda_1 \lambda_2}(\bphat_1, \bkhat) G_{\rho\sigma;\lambda_1 \lambda_2}(\bkhat, \bphat_3)\notag\\
    &\qquad\qquad \times \bar{u}_{\lambda_3}(p_3) \gamma^\rho \left(\gamma^0 + \hat{k}_j \gamma^j \right) \gamma^\mu u_{\lambda_1}(p_1)\notag\\
    &\qquad\qquad \times \bar{u}_{\lambda_4}(p_4) \gamma^{\sigma}\left( \gamma^0 - \hat{k}_l \gamma^l\right) \gamma^\nu u_{\lambda_2}(p_2)\Big\}\,.
    \tag{\ref{eq:calT}}
\end{align}

Let us first focus on the integral $\mathcal{I}$~\eqref{eq:calI}.
We first evaluate the $k_0$-integral by performing the Wick rotation $k_0 \to i k_4$:
\begin{align}
    &i\int_{-\infty}^{\infty}\frac{d k_4}{2\pi}  \, 
    \frac{1}{\left(ik_4 + \epsilon_k \right) \left(ik_4 - \epsilon_k\right)} = -\frac{i}{2 |\epsilon_k|}\,.
\end{align}
Then, by performing the change of variables $k \to \epsilon_k = k - \mu$, we can perform the $k$-integral as
\begin{align}
    \mathcal{I} &= \frac{iN}{8} \int_{e^{-t}\Lambda < |\epsilon_k| < \Lambda} d\epsilon_k \, \frac{1}{|\epsilon_k|}\,,\notag\\
    &= \frac{i N}{4} t\,,
\end{align}
where the factor two comes from the combination of positive and negative integration ranges of $\epsilon_k$.

Next, we evaluate the gamma matrix algebra in $\mathcal{T}_{\lambda_1\lambda_2}$~\eqref{eq:calT}.
By substituting the actual expression for $G_{\mu\nu}$~\eqref{eq:Gmunu}, we obtain
\begin{align}
    \mathcal{T}_{\lambda_1 \lambda_2}
    &=\int \frac{d\hat{k}}{4\pi} \bigg\{ G^0_{\lambda_1 \lambda_2}(\bphat_1, \bkhat) G^0_{\lambda_1 \lambda_2}(\bkhat,\bphat_3) \bar{u}_{\lambda_3} \gamma^0 \gamma^0 \gamma^0 u_{\lambda_1} \bar{u}_{\lambda_4} \gamma^0 \gamma^0 \gamma^0 u_{\lambda_2}\notag \\
    & \qquad\qquad\ - \bigg[\frac13 G^0_{\lambda_1 \lambda_2}(\bphat_1,\bkhat) G^0_{\lambda_1 \lambda_2}(\bkhat,\bphat_3) + G^0_{\lambda_1 \lambda_2}(\bphat_1, \bkhat) G^i_{\lambda_1 \lambda_2}(\bkhat, \bphat_3) \notag\\
    &\qquad\qquad\qquad + G^i_{\lambda_1 \lambda_2}(\bphat_1,\bkhat) G^0_{\lambda_1 \lambda_2}(\bkhat,\bphat_3) \bigg] \bar{u}_{\lambda_3} \gamma^i \gamma^0 \gamma^0 u_{\lambda_1} \bar{u}_{\lambda_4} \gamma^i \gamma^0 \gamma^0 u_{\lambda_2} \notag\\
    & \qquad \qquad \ + \bigg[\frac13 G^0_{\lambda_1 \lambda_2}(\bphat_1, \bkhat) G^i_{\lambda_1 \lambda_2}(\bkhat, \bphat_3) + \frac13 G^i_{\lambda_1 \lambda_2}(\bphat_1,\bkhat) G^0_{\lambda_1 \lambda_2}(\bkhat,\bphat_3)\notag\\
    &\qquad\qquad\qquad + G^i_{\lambda_1 \lambda_2}(\bphat_1,\bkhat) G^i_{\lambda_1 \lambda_2}(\bkhat,\bphat_3) \bigg] \bar{u}_{\lambda_3} \gamma^i \gamma^j \gamma^0 u_{\lambda_1} \bar{u}_{\lambda_4} \gamma^i \gamma^j \gamma^0 u_{\lambda_2} \notag\\
    & \qquad \qquad \ - \frac13 G^i_{\lambda_1 \lambda_2}(\bphat_1,\bkhat) G^i_{\lambda_1 \lambda_2}(\bkhat,\bphat_3) \bar{u}_{\lambda_3} \gamma^i \gamma^j \gamma^k u_{\lambda_1} \bar{u}_{\lambda_4} \gamma^i \gamma^j \gamma^k u_{\lambda_2} \bigg\} \,,
    \label{eq:calT12}
\end{align}
where we used the property $\int \frac{d\hat{k}}{4\pi} \, \hat{k}_i = 0$ and $\int \frac{d\hat{k}}{4\pi}\,  \hat{k}_i \hat{k}_j = \delta_{ij}/3$.
The helicity conservation requires $\lambda_1 = \lambda_3$ and $\lambda_2 = \lambda_4$.
For the helicity $\lambda_1 = +$ and $\lambda_2 = \pm$ channel (the double signs are in the same order below):
\begin{align}
    \bar{u}_{+}\gamma^0 \gamma^0 \gamma^0 u_{+} \bar{u}_{\pm} \gamma^0 \gamma^0 \gamma^0 u_{\pm}&= \bar{u}_{+}\gamma^0 u_{+} \bar{u}_{\pm} \gamma^0 u_{\pm}\,,\\
    \bar{u}_{+}\gamma^i \gamma^0 \gamma^0 u_{+} \bar{u}_{\pm} \gamma^i \gamma^0 \gamma^0 u_{\pm}&= \bar{u}_{+}\gamma^i u_{+} \bar{u}_{\pm} \gamma^i u_{\pm}\,,\\
    \bar{u}_{+}\gamma^i \gamma^j \gamma^0 u_{+} \bar{u}_{\pm} \gamma^i \gamma^j \gamma^0 u_{\pm}&= 3\bar{u}_{+}\gamma^0 u_{+} \bar{u}_{\pm} \gamma^0 u_{\pm} \mp 2\bar{u}_{+}\gamma^i u_{+} \bar{u}_{\pm} \gamma^i u_{\pm}\,,\\
    \bar{u}_{+}\gamma^i \gamma^j \gamma^k u_{+} \bar{u}_{\pm} \gamma^i \gamma^j \gamma^k u_{\pm}&= \mp 6\bar{u}_{+}\gamma^0 u_{+} \bar{u}_{\pm} \gamma^0 u_{\pm} + 7\bar{u}_{+}\gamma^i u_{+} \bar{u}_{\pm} \gamma^i u_{\pm}\,,
\end{align}

In the helicity $++$ channel, Eq.~\eqref{eq:calT12} simplifies to
\begin{align}
    \mathcal{T}_{++}
    &=\int \frac{d\hat{k}}{4\pi}
    \bigg\{
        \Big[G^0_{++}(\bphat_1,\bkhat) G^0_{++}(\bkhat,\bphat_3) + G^0_{++}(\bphat_1, \bkhat) G^i_{++}(\bkhat, \bphat_3) \\
        &\qquad\qquad\quad + G^i_{++}(\bphat_1,\bkhat) G^0_{++}(\bkhat,\bphat_3) + 5 G^i_{++}(\bphat_1,\bkhat) G^i_{++}(\bkhat,\bphat_3) \Big] \bar{u}_{+} \gamma^0 u_{+} \bar{u}_{+} \gamma^0 u_{+} \notag\\
        &\qquad\qquad\ -\frac13 \Big[G^0_{++}(\bphat_1,\bkhat) G^0_{++}(\bkhat,\bphat_3) + 5 G^0_{++}(\bphat_1, \bkhat) G^i_{++}(\bkhat, \bphat_3) \notag\\
        &\qquad\qquad\qquad\ + 5 G^i_{++}(\bphat_1,\bkhat) G^0_{++}(\bkhat,\bphat_3) + 13 G^i_{++}(\bphat_1,\bkhat) G^i_{++}(\bkhat,\bphat_3) \Big] \bar{u}_{+} \gamma^i u_{+} \bar{u}_{+} \gamma^i u_{+}
    \bigg\} \,, \notag
\end{align}
and in the helicity $+-$ channel, Eq.~\eqref{eq:calT12} simplifies to
\begin{align}
    \mathcal{T}_{+-}
    &=\int \frac{d\hat{k}}{4\pi}
    \bigg\{
        \Big[G^0_{+-}(\bphat_1,\bkhat) G^0_{+-}(\bkhat,\bphat_3) + G^0_{+-}(\bphat_1, \bkhat) G^i_{+-}(\bkhat, \bphat_3) \\
        &\qquad\qquad\quad + G^i_{+-}(\bphat_1,\bkhat) G^0_{+-}(\bkhat,\bphat_3) + G^i_{+-}(\bphat_1,\bkhat) G^i_{+-}(\bkhat,\bphat_3) \Big] \bar{u}_{+} \gamma^0 u_{+} \bar{u}_{-} \gamma^0 u_{-} \notag\\
        &\qquad\qquad\ -\frac13 \Big[G^0_{+-}(\bphat_1,\bkhat) G^0_{+-}(\bkhat,\bphat_3) +  G^0_{+-}(\bphat_1, \bkhat) G^i_{+-}(\bkhat, \bphat_3) \notag\\
        &\qquad\qquad\qquad\ +  G^i_{+-}(\bphat_1,\bkhat) G^0_{+-}(\bkhat,\bphat_3) +  G^i_{+-}(\bphat_1,\bkhat) G^i_{+-}(\bkhat,\bphat_3) \Big] \bar{u}_{+} \gamma^i u_{+} \bar{u}_{-} \gamma^i u_{-}
    \bigg\} \,. \notag
\end{align}
Now, we perform the partial wave expansion in the above equations:
\begin{equation}
    G(\bphat, \bkhat) = \sum_L (2L+1) G_L P_L(\bphat \cdot \bkhat)\,.
\end{equation}
We express the amplitude in terms of the coupling function in the orbital angular momentum channel $L$.
The component of $\mathcal{T}_{++}$ in the $L$-channel is
\begin{align}
    \mathcal{T}_{L++}
    &= \left[\left(G^0_{L++}\right)^2 +  2 G^0_{L++} G^i_{L++} + 5\left(G^i_{L++}\right)^2 \right] \bar{u}_{+} \gamma^0 u_{+} \bar{u}_{+} \gamma^0 u_{+}\notag \\
    & \quad - \frac13\left[
    \left(G^0_{L++}\right)^2 +10 G^0_{L++} G^i_{L++} + 13 \left(G^i_{L++}\right)^2 \right] \bar{u}_{+} \gamma^i u_{+} \bar{u}_{+} \gamma^i u_{+}\,,    
\end{align}
and for $\mathcal{T}_{+-}$ in the $L$-channel, it is
\begin{align}
    \mathcal{T}_{L+-}
    &= \left[\left(G^0_{L+-}\right)^2 +  2 G^0_{L+-} G^i_{L+-} + \left(G^i_{L+-}\right)^2 \right] \bar{u}_{+} \gamma^0 u_{+} \bar{u}_{-} \gamma^0 u_{-} \notag \\
    & \quad - \frac13 \left[
    \left(G^0_{L+-}\right)^2 + 2 G^0_{L+-} G^i_{L+-} + \left(G^i_{L+-}\right)^2 \right] \bar{u}_{+} \gamma^i u_{+} \bar{u}_{-} \gamma^i u_{-}\,,    
\end{align}
By comparing Eq.~\eqref{eq:deltaM} with Eq.~\eqref{eq:deltaMIT}, one obtains the one-loop correction to the coupling functions as
\begin{align}
    \delta G^0_{L++} &= -\frac{N}4 t \left[\left(G^0_{L++}\right)^2 +  2 G^0_{L++} G^i_{L++} + 5\left(G^i_{L++}\right)^2 \right]\,, \\
    \delta G^i_{L++} &= - \frac{N}{12}t \left[
    \left(G^0_{L++}\right)^2 +10 G^0_{L++} G^i_{L++} + 13 \left(G^i_{L++}\right)^2 \right] \,, \\
    \delta G^0_{L+-} &= -\frac{N}4 t \left[\left(G^0_{L+-}\right)^2 +  2 G^0_{L+-} G^i_{L+-} + \left(G^i_{L+-}\right)^2 \right]\,, \\
    \delta G^i_{L+-} &= - \frac{N}{12}t \left[
    \left(G^0_{L+-}\right)^2 + 2G^0_{L+-} G^i_{L+-} + \left(G^i_{L+-}\right)^2 \right] \,.   
\end{align}
Then, the RG equation from the one-loop correction is
\begin{align}
    \frac{d G_{L\lambda_1 \lambda_2}^{0/i}}{dt} = \delta \beta^{0/i}_{L\lambda_1 \lambda_2}\,,
\end{align}
where the one-loop beta function associated with the one-loop correction to the coupling function is
\begin{align}
    \delta \beta^{0/i}_{L\lambda_1 \lambda_2} = \frac{d\delta G^{0/i}_{L\lambda_1 \lambda_2}}{dt}\,.
\end{align}
This completes the derivation of the RG equations (\ref{eq:RG0++}, \ref{eq:RGi++}, \ref{eq:RG0+-}, \ref{eq:RGi+-}).

\subsection{Solution of the RG equation in perturbation theory}
\label{sec:solRGeq}

Let us discuss the solution of the differential equation \eqref{eq:generalRG} and obtain the solution in perturbation theory up to the next-to-leading order in perturbation theory.
The differential equation we consider is
\begin{equation}
    \frac{d y(t)}{dt} = - \frac{a}{h^2(t)} y^2(t) - \frac{c^2}{4a} \gbar^2\,,
    \tag{\ref{eq:generalRG}}
\end{equation}
The solution for the initial condition $y(t=0) = - b g^2 l / \pi^2$ is
\begin{equation}
    y(t) = -\frac{c \gbar h(t) \mathcal{X}}{2a \mathcal{Y}}\,,
    \tag{\ref{eq:generalsol}}
\end{equation}
where the numerator and denominator is given by
\begin{align}
    \tag{\ref{eq:calX}}
    \mathcal{X}
    &= \left[c\gbar l J_0\left(\frac{1}{\gbar}\right) - 2 J_1\left(\frac{1}{\gbar}\right) \right] Y_1\left(\frac{h(t)}{\gbar}\right)
    - \left[c\gbar l Y_0\left(\frac{1}{\gbar}\right) - 2 Y_1\left(\frac{1}{\gbar}\right) \right] J_1\left(\frac{h(t)}{\gbar}\right)\,,\\
    \mathcal{Y}
    &= \left[c\gbar l J_0\left(\frac{1}{\gbar}\right) - 2 J_1\left(\frac{1}{\gbar}\right) \right] Y_0\left(\frac{h(t)}{\gbar}\right)
    - \left[c\gbar l Y_0\left(\frac{1}{\gbar}\right) - 2 Y_1\left(\frac{1}{\gbar}\right) \right] J_0\left(\frac{h(t)}{\gbar}\right)\,.
    \tag{\ref{eq:calY}}
\end{align}
Now we expand these Bessel functions in terms of $\gbar \ll 1$, using the asymptotic formulas (\ref{eq:BesselJ}, \ref{eq:BesselY}).
We can approximate as
\begin{align}
    \label{eq:J01g}
    J_0\left(\frac{1}{\gbar}\right) &\simeq \frac{\sqrt{\gbar}}{8\sqrt{\pi}} \left[(8-\gbar) \cos \left(\frac{1}{\gbar}\right) + (8 + \gbar) \sin \left(\frac{1}{\gbar}\right)\right]\,,\\
    J_1\left(\frac{1}{\gbar}\right) &\simeq \frac{\sqrt{\gbar}}{8\sqrt{\pi}} \left[(-8+3\gbar) \cos \left(\frac{1}{\gbar}\right) + (8 + 3\gbar) \sin \left(\frac{1}{\gbar}\right)\right]\,,\\
    Y_0\left(\frac{1}{\gbar}\right) &\simeq - \frac{\sqrt{\gbar}}{8\sqrt{\pi}} \left[(8+\gbar) \cos \left(\frac{1}{\gbar}\right) + (-8 + \gbar) \sin \left(\frac{1}{\gbar}\right)\right]\,,\\
    Y_1\left(\frac{1}{\gbar}\right) &\simeq - \frac{\sqrt{\gbar}}{8\sqrt{\pi}} \left[(8 + 3\gbar) \cos \left(\frac{1}{\gbar}\right) + (8 - 3 \gbar) \sin \left(\frac{1}{\gbar}\right)\right]\,.
\end{align}
Also, because $h(t) / \gbar \simeq 1/\gbar + c\gbar t /2 \gg 1$, the following expansion is valid:
\begin{align}
    J_0\left(\frac{h(t)}{\gbar}\right) &\simeq \frac{\sqrt{\gbar}}{8\sqrt{\pi}} \left[(8-\gbar) \cos \left(\frac{h(t)}{\gbar}\right) + (8 + \gbar) \sin \left(\frac{h(t)}{\gbar}\right)\right]\,,\\
    J_1\left(\frac{h(t)}{\gbar}\right) &\simeq \frac{\sqrt{\gbar}}{8\sqrt{\pi}} \left[(-8+3\gbar) \cos \left(\frac{h(t)}{\gbar}\right) + (8 + 3\gbar) \sin \left(\frac{h(t)}{\gbar}\right)\right]\,,\\
    Y_0\left(\frac{h(t)}{\gbar}\right) &\simeq -\frac{\sqrt{\gbar}}{8\sqrt{\pi}} \left[(8+\gbar) \cos \left(\frac{h(t)}{\gbar}\right) + (-8 + \gbar) \sin \left(\frac{h(t)}{\gbar}\right)\right]\,,\\
    Y_1\left(\frac{h(t)}{\gbar}\right) &\simeq -\frac{\sqrt{\gbar}}{8\sqrt{\pi}} \left[(8+3\gbar) \cos \left(\frac{h(t)}{\gbar}\right) + (8 - 3\gbar) \sin \left(\frac{h(t)}{\gbar}\right)\right]\,.
\end{align}
We further expand trigonometric functions in the above equations as
\begin{align}
    \cos\left(\frac{h(t)}{\gbar}\right) &= \cos \left(\frac{1}{\gbar} + \frac{h(t)-1}{\gbar}\right) \notag\\
    &= \cos \left(\frac{1}{\gbar}\right) \cos\left(\frac{h(t) - 1}{\gbar}\right) - \sin\left(\frac{1}{\gbar}\right) \sin \left(\frac{h(t)-1}{\gbar}\right)\,, \\
    \sin\left(\frac{h(t)}{\gbar}\right) &= \sin \left(\frac{1}{\gbar} + \frac{h(t)-1}{\gbar}\right) \notag\\
    &= \sin \left(\frac{1}{\gbar}\right) \cos\left(\frac{h(t) - 1}{\gbar}\right) + \cos\left(\frac{1}{\gbar}\right) \sin \left(\frac{h(t)-1}{\gbar}\right)\,.
    \label{eq:sinhg}
\end{align}
Then, by substituting Eqs.~(\ref{eq:J01g}--\ref{eq:sinhg}) into Eqs.~(\ref{eq:calX}, \ref{eq:calY}), $\sin(1/\gbar)$ and $\cos(1/\gbar)$ disappear from the final expression owing to the relation $\sin^2(1/\gbar) + \cos^2(1/\gbar) = 1$, and $\mathcal{X}$ and $\mathcal{Y}$ become
\begin{align}
    \mathcal{X} &\simeq \left(64 - 3\gbar^2\right) c\gbar l  \cos\left(\frac{h(t) - 1}{\gbar}\right) + 2 \left[64 + \left(9 - 16 cl \right)\gbar^2 \right] \sin\left(\frac{h(t) - 1}{\gbar}\right)\,,\\
    \mathcal{Y} &\simeq 2 \left(64 - 3\gbar^2\right) \cos\left(\frac{h(t) - 1}{\gbar}\right) + \left[64 - \left(64 + \gbar^2\right)c l \right] \gbar \sin\left(\frac{h(t) - 1}{\gbar}\right)\,.
\end{align}
By further substituting these into Eq.~\eqref{eq:generalsol}, we get the final solution up to the next-to-leading order in perturbation theory as
\begin{align}
    y(t) &\simeq - \frac{c \gbar \sqrt{1 + c \gbar^2 t}}{2a}
    \frac{\left(64 - 3\gbar^2\right) c\gbar l  + 2 \left[64 + \left(9 - 16 cl \right)\gbar^2 \right] \tan\left(\frac{h(t) - 1}{\gbar}\right)}
    {2 \left(64 - 3\gbar^2\right) + \left[64 - \left(64 + \gbar^2\right)c l \right] \gbar \tan\left(\frac{h(t) - 1}{\gbar}\right)}\notag\\
    &\simeq
    - \frac{c \gbar}{2a}
    \frac{c\gbar l  + 2 \tan\left(\frac{h(t) - 1}{\gbar}\right)}
    {2 + \left(1 - cl\right)\gbar \tan\left(\frac{h(t) - 1}{\gbar}\right)}\,.
\end{align}
From this final expression, Eq.~\eqref{eq:tanBCS} can be obtained by setting the denominator equals to zero.

\bibliography{pairing}

\end{document}